\documentclass[a4paper,11pt]{article}

\makeatletter
\def\@seccntformat#1{\@ifundefined{#1@cntformat}%
   {\csname the#1\endcsname\quad}  
   {\csname #1@cntformat\endcsname}
}
\let\oldappendix\appendix 
\renewcommand\appendix{%
    \oldappendix
    \newcommand{\section@cntformat}{\appendixname~\thesection:\,\,}
}
\makeatother

\usepackage{geometry}
\usepackage[utf8]{inputenc}
\usepackage[hidelinks]{hyperref}
\usepackage{natbib}
\usepackage{amsfonts}
\usepackage{amsmath}
\usepackage{xcolor}
\usepackage{graphicx}
\usepackage{placeins}
\usepackage{float}
\usepackage{enumitem}
\usepackage{subfigure}
\usepackage{tabularx}
\usepackage{xurl}
\usepackage{multicol}
\usepackage{multirow}
\usepackage{colortbl}
\usepackage{booktabs}
\usepackage{physics}
\usepackage{xltabular}
\usepackage[font=footnotesize,skip=7pt]{caption}
\usepackage[normalem]{ulem} 

\usepackage{longtable}
\usepackage{array}

\newtheorem{definition}{Definition}

\definecolor{armygreen}{rgb}{0.19, 0.53, 0.43}

\title{Rank-size analysis of optimal portfolio weights with a reverse-engineering application}

\author{%
  Roy Cerqueti\textsuperscript{1,2}\quad
  Francesco Cesarone\textsuperscript{3,*}\quad
  Alessio Di Paolo\textsuperscript{3}\quad
  Valerio Ficcadenti\textsuperscript{4}%
}
\date{}

\begin{document}

\maketitle

\begin{center}
\small
\textsuperscript{1}Department of Social and Economic Sciences,
Sapienza University of Rome,\\
Piazzale Aldo Moro 5, 00185 Rome, Italy\par

\textsuperscript{2}GRANEM, University of Angers,
SFR CONFLUENCES, F-49000 Angers, France\par

\textsuperscript{3}Department of Business Studies, Roma Tre University,\\
Via Silvio D'Amico 77, 00145 Rome, Italy\par

\textsuperscript{4}LSBU Business School, London South Bank University,\\
103 Borough Road, London SE1 0AA, United Kingdom\par

\bigskip
\textsuperscript{*}\textbf{Corresponding author: Francesco Cesarone}\par
\bigskip


\begin{tabular}{@{}lll@{}}
\textbf{Author} & \textbf{Email} & \textbf{ORCID} \\
Roy Cerqueti &
  \href{mailto:roy.cerqueti@uniroma1.it}{roy.cerqueti@uniroma1.it} &
  \href{https://orcid.org/0000-0002-1871-7371}{0000-0002-1871-7371} \\
Francesco Cesarone &
  \href{mailto:francesco.cesarone@uniroma3.it}{francesco.cesarone@uniroma3.it} &
  \href{https://orcid.org/0000-0003-2326-4204}{0000-0003-2326-4204} \\
Alessio Di Paolo &
  \href{mailto:alessio.dipaolo@uniroma3.it}{alessio.dipaolo@uniroma3.it} &
  \href{https://orcid.org/0009-0006-7469-6073}{0009-0006-7469-6073} \\
Valerio Ficcadenti &
  \href{mailto:ficcadv2@lsbu.ac.uk}{ficcadv2@lsbu.ac.uk} &
  \href{https://orcid.org/0000-0001-8786-8213}{0000-0001-8786-8213}
\end{tabular}
\end{center}


\begin{abstract}
This paper investigates the shape of optimal portfolio weights using rank-size
analysis. 
We consider four portfolio selection models: Global Minimum-Variance (GMV), minimum-CVaR, Risk Parity (RP), and the Most Diversified Portfolio (MDP). For each model, we fit seven rank-size
specifications to the sorted portfolio weights generated over rolling windows. 
To compare the fitted specifications, we combine standard goodness-of-fit measures with Cumulative Second-order $\varepsilon$-Stochastic Dominance, an $\varepsilon$-parameterized generalization of Second-order Stochastic Dominance that quantifies the extent to which the exact dominance conditions are satisfied or violated.
Empirically applied to major equity indices, this approach provides a systematic criterion for identifying the optimal functional form across diverse market compositions. Furthermore, we address the inverse problem of identifying the generating model from the observed weight structure. We use the calibrated rank-size
parameters as inputs to a Random Forest classifier, evaluated by a blocked
cross-validation that accounts for the overlap between estimation windows.
The parameters separate minimum-risk from risk-diversification strategies and identify the RP and MDP portfolios, whereas the GMV and minimum-CVaR portfolios are difficult to distinguish. These results indicate that rank-size parameters provide a compact and informative description of
portfolio allocation geometries.

\medskip

\noindent
\textbf{Keywords}: Portfolio optimization; Rank-size analysis; Stochastic dominance; Reverse engineering; Machine learning.
\end{abstract}

\newpage

\section{Introduction}

In portfolio management, optimal asset allocation plays a key role in achieving desired investment objectives in terms of risk-return. Traditional portfolio optimization models, such as Mean-Variance (MV) portfolio selection, pioneered by \cite{Markowitz1952, Markowitz1959}, have laid the foundation of modern portfolio theory by offering a systematic approach to asset allocation.
However, the increasing complexity of financial markets and the diversity of investor preferences highlight the need for more advanced models.
These include the minimum Conditional Value-at-Risk (CVaR) portfolio \citep{rockafellar2000optimization}, the Risk Parity (RP) portfolio \citep{maillard2010properties}, and the Most Diversified Portfolio (MDP) \citep{choueifaty2008towards}, each offering a distinct approach to risk management and diversification.
In this context, attention has predominantly focused on optimal objective values and out-of-sample performance \citep[see, e.g.,][]{jagannathan2003risk, DeMiguel2009, bruni2017exact}, rather than on the structural analysis of the weights themselves.
Concurrently, a rich literature has shown that heavy-tailed returns and model uncertainty may substantially affect diversification benefits and the behavior of risk measures, further complicating the analysis of portfolio construction rules \citep[see, e.g.,][]{ibragimov2008portfolio,Ibragimov01082009}.
Two portfolios can have similar aggregate risk and yet allocate capital very
differently. One portfolio may be dominated by a few large positions, whereas
another may distribute capital more gradually across assets. These differences
are not captured by the objective value alone, but they provide information
about concentration, diversification, and the portfolio construction rule.

\noindent
In this paper, we study this structure by sorting the optimal weights in decreasing order and
analyzing how they decay with rank. We refer to the resulting curve as the
``geometry'' of the portfolio weights. It describes how capital is distributed
between the largest positions and the tail of the portfolio and provides a
common representation across models, markets, and time.

Rank-size analysis describes the relationship between the ordered values of a
variable and their ranks \citep{naumis2008tail,ausloos2016universal}. Its
parameters provide a compact description of the concentration, decay rate, and
tail of the ranked distribution. This makes rank-size analysis suitable for
comparing portfolio weight structures with different numbers of assets and
different allocation patterns.

To achieve a comprehensive representation of the weight allocation structure, we apply various rank-size functions (Exponential, Power, Shifted Power, Discrete Generalized Beta Distribution, Zipf--Mandelbrot, Polynomial, and Universal laws) to optimal weights generated by different portfolio selection models (GMV, CVaR, RP, MDP).
Empirically, this analysis is conducted on four major asset universes (Dow Jones, EURO STOXX 50, NASDAQ 100, FTSE 250) and adopts a rolling time window approach to ensure robustness.
This procedure yields a specific set of estimated parameters that characterize how capital is distributed across assets and serves as a tool to extract valuable information on optimization strategies and models.
Comparative analyses of rank-size parameters have been widely adopted across diverse domains, ranging from urban dynamics \citep{giesen2011zipf} and linguistics to sports analytics \citep{ficcadenti2023rank}, demonstrating the robustness of this approach in capturing structural characteristics.
In the specific context of financial markets, \citet{campbell2025macroscopic} recently analyzed the stability of the capital distribution curve, showing that the cross-sectional geometry of market weights has a direct impact on portfolio performance.
Although our focus shifts to the geometry of optimal portfolio weights generated by various optimization models, the underlying idea of summarizing cross-sectional structure via a smooth curve remains closely related.

However, applying this framework introduces a critical methodological challenge: selecting the ``best'' functional form to represent the data.
In the existing literature, selection practices vary significantly: some rely on graphical inspection and simple goodness-of-fit measures such as the coefficient of determination ($R^2$) \citep[see, e.g.,][]{axtell2001zipf}, while others employ error-based criteria like Root Mean Squared Error (RMSE) or Mean Absolute Error (MAE) \citep[see, e.g.,][]{giesen2010size}.
Consequently, there is no universally agreed standard for selecting among competing rank-size laws, especially when the goal is to compare performance across many realisations of the same phenomenon \citep[see, e.g.,][]{CERQUETI2022111975}.
This heterogeneity complicates cross-study comparisons, motivating the development of more systematic selection procedures.
In this paper, we address this issue by integrating standard goodness-of-fit metrics with a stochastic dominance criterion on the error distributions.
This approach enables us to robustly identify the rank-size law that best characterizes the optimal portfolio weight distribution.

This paper makes three main contributions. 
\textit{First}, we provide a common rank-size representation of the optimal weights generated by four portfolio selection models, allowing their concentration and decay patterns to be compared across markets and rolling windows. 
\textit{Second}, we propose a selection procedure that compares the entire empirical distributions of the fitting errors through Cumulative Second-order-$\varepsilon$-Stochastic Dominance (CS$\varepsilon$SD), a generalization of stochastic dominance depending on a parameter $\varepsilon$, which is particularly appropriate in our context.
\textit{Third}, we assess, under a validation design that accounts for the overlap between estimation windows, whether the parameters of the selected rank-size law can distinguish between minimum-risk and risk-diversification
strategies and identify the specific portfolio model that generated the weights.

The remainder of the paper is organized as follows.
Sect.~\ref{Sec:literature} reviews the related literature.
Sects.~\ref{Sec:Portfolio_sel}, \ref{Sec:ranksize}, \ref{Sec:evaluation}, and \ref{Sec:reverse} present the portfolio models,
rank-size specifications, selection procedure, and classification framework, respectively.
Sect.~\ref{Sec:empanalysis} reports the empirical results, while
Sect.~\ref{Sec:conclusions} concludes.

\section{Literature review} \label{Sec:literature}

The classical MV portfolio selection model \citep{Markowitz1952, Markowitz1959} is widely recognized as the cornerstone in modern portfolio theory. It provides a systematic framework for asset allocation by minimizing portfolio variance for a given level of expected return, effectively balancing risk and return.
Despite its foundational importance and widespread use among both academics and practitioners, the MV model has faced criticism and motivated the development of alternative models that incorporate different risk measures \citep[see, e.g.,][]{king1993asymmetric,konno1991mean, mitra2003review,rockafellar2000optimization, young1998minimax}.
However, minimum-risk strategies, while effective  at reducing overall risk, can lead to portfolios that are poorly diversified in terms of risk allocation \citep{maillard2010properties, roncalli2013introduction}. This insufficient diversification may lead to portfolios that are overly concentrated in a few assets, exposing investors to idiosyncratic risks.
A systematic comparison of minimum-risk, capital-diversification, and
risk-diversification strategies is provided by
\citet{cesarone2018minimum}, who also extend the equal-risk-contribution
framework to CVaR.
In response to these challenges, particularly those highlighted during the 2008 global financial crisis, portfolio strategies that emphasize risk diversification have gained prominence. A simple example is the Equally Weighted (EW) portfolio, in which capital is distributed equally among all assets. While this approach maximizes allocation diversification, it ignores differences in asset risk profiles and correlations, which can result in suboptimal risk-adjusted returns.
Other approaches focus on risk allocation. 
Among the most prominent, both theoretically and practically, are the Risk Parity (RP) portfolio \citep{maillard2010properties,roncalli2013introduction} and the Most Diversified Portfolio (MDP) \citep{choueifaty2008towards,choueifaty2013properties}. 
The RP approach aims to select a portfolio in which each asset contributes equally to the total portfolio risk. This is achieved by quantifying each asset’s risk contribution using Euler's decomposition of a homogeneous risk function. The MDP, on the other hand, is obtained by maximizing the Diversification Ratio, defined as the ratio of the weighted average asset volatility to the portfolio volatility. 
Both strategies typically use volatility as the risk metric and are collectively referred to as \textit{risk-diversification} strategies.
Although the two constructions are generally distinct,
\citet{cesarone2025return} show that volatility-based RP and MDP portfolios
coincide under equicorrelated asset returns and extend the
diversification-ratio approach to general risk measures.
Several other diversification approaches have also been proposed. These include entropy-based methods \citep[see, e.g.,][]{bera2008optimal, usta2011mean,pola2016entropy}, extensions of risk parity that incorporate general risk measures \citep[see, e.g.,][]{ararat2024mad,bai2016least,cesarone2017equal,chaves2012efficient,bellini2021risk}, and the Minimum-Torsion Bets approach \citep{meucci2015risk}.
Because these portfolio models pursue different objectives, they may generate
weight profiles with different levels of concentration and different decay
patterns. A common representation of these profiles is therefore needed to
compare how the models distribute capital across assets.

Optimal portfolio weights have also been studied as random objects in their own right. In the classical Mean--Variance framework, \citet{OKHRIN2006235} show that estimated optimal weights have non-trivial finite-sample distributions, often displaying skewness, heavy tails and, in some cases, undefined moments, particularly for Sharpe-ratio-based portfolios. Their analysis highlights the extent to which optimal weights can vary across samples and how sensitive they are to small perturbations in means and covariances. This distributional complexity motivates approaches that do not rely on strong parametric assumptions about the weights, but instead aim to describe their empirical shapes in a model-agnostic way.
A complementary perspective is provided by the literature on diversification under heavy-tailed risks. The theoretical insights of \cite{ibragimov2008portfolio}, \cite{ Ibragimov01082009}, and related works demonstrate that when returns exhibit heavy-tailed or Pareto-type behavior, diversification may fail to deliver the conventional smoothing effects, and portfolios can remain inherently concentrated even before optimization is applied. In such environments, optimization rules interact with the tail properties of returns to produce weight structures that may be far from the Gaussian benchmark. 
This mechanism is consistent with the empirical patterns we document: different optimization models tend to accentuate or moderate these tail-driven forces, leading to distinct cross-sectional shapes in the distribution of optimal weights.

Existing research does not provide a common functional representation for
comparing optimal weight profiles across portfolio models. Rank-size analysis
offers a natural approach because it models how an ordered quantity decreases
with rank and summarizes this pattern through a small number of parameters
\citep{naumis2008tail,ausloos2016universal}. Several specifications have been
proposed, including exponential, power-law, shifted power-law, Discrete Generalized Beta Distribution (DGBD),
Zipf--Mandelbrot, polynomial, and universal forms. Their definitions and parameter
interpretations are presented in Sect.~\ref{Sec:ranksize}.

A related application is provided by \citet{dombrowski2023rank}, who use
rank-size specifications to study commercial real-estate transaction volumes
and interpret the associated probability masses as portfolio weights. However,
we are not aware of a study that systematically compares alternative rank-size
laws for weights generated by different portfolio optimization models or uses
their fitted parameters to classify the generating model. This is the gap
addressed in this paper.

%
\section{Methodology}

The methodology consists of four stages. First, we generate optimal portfolio
weights using the GMV, minimum-CVaR, RP, and MDP models
(Sect.~\ref{Sec:Portfolio_sel}). Second, we sort the weights and fit the
candidate rank-size specifications (Sect.~\ref{Sec:ranksize}). Third, we
compare the fitted specifications using standard goodness-of-fit measures and
stochastic dominance applied to their error distributions
(Sect.~\ref{Sec:evaluation}). Finally, we use the parameters of the selected
specification as inputs to a Random Forest classifier and assess whether they
identify the portfolio strategy and model (Sect.~\ref{Sec:reverse}).

\subsection{Portfolio selection models} \label{Sec:Portfolio_sel}

We consider an investment universe of $n$ assets, whose linear returns are
represented by the random variables $R_1,\ldots,R_n$. Under full-investment
and long-only constraints, a portfolio is identified by a vector
$\boldsymbol{x}=(x_1,\ldots,x_n)^{\top}$ belonging to the standard simplex
$
\Delta
=
\left\{
\boldsymbol{x}\in\mathbb{R}^{n}:
\sum_{k=1}^{n}x_k=1,\quad
x_k\geq 0,\quad k=1,\ldots,n
\right\},
$
where $x_k$ denotes the fraction of capital invested in asset $k$. The
portfolio return is
$
R_P(\boldsymbol{x})
=
\sum_{k=1}^{n}x_kR_k.
$

\noindent
We assume that asset returns are represented by $T$ equally likely historical
scenarios. For each scenario $t=1,\ldots,T$, let $r_{k,t}$ denote the return
of asset $k$. The corresponding portfolio return realization is
$
R_{P,t}(\boldsymbol{x})
=
\sum_{k=1}^{n}x_kr_{k,t}.
$

We consider four portfolio selection models. 
The Global Minimum-Variance (GMV) and minimum-CVaR models belong to the class of minimum-risk strategies, whereas the Risk Parity (RP) and Most Diversified Portfolio (MDP) models belong to the class of risk-diversification strategies. 
Since these models
pursue different objectives, they may generate substantially different
portfolio weight structures.

%
\subsubsection{Minimum-Risk strategies}\label{Sec:minrisk}

Minimum-risk strategies select a portfolio that minimizes a given risk measure
over the feasible set $\Delta$. Since different risk measures capture different
features of the portfolio-return distribution, they may produce substantially
different optimal allocations. In this paper, we consider portfolio variance
and CVaR.

\paragraph{Global minimum-variance portfolio.}

Let $\boldsymbol{\Sigma}=(\sigma_{k\ell})_{k,\ell=1}^{n}$ denote the
covariance matrix of asset returns. The portfolio variance is
\begin{equation*}
\sigma_P^2(\boldsymbol{x})
=
\boldsymbol{x}^{\top}\boldsymbol{\Sigma}\boldsymbol{x}
=
\sum_{k=1}^{n}\sum_{\ell=1}^{n}x_kx_\ell\sigma_{k\ell}.
\end{equation*}
The GMV portfolio is obtained by solving the following convex quadratic
programming problem:
\begin{equation}
\label{eq:gmv}
\begin{aligned}
\boldsymbol{x}^{\mathrm{GMV}}
\in\arg\min_{\boldsymbol{x}}\quad
& \boldsymbol{x}^{\top}\boldsymbol{\Sigma}\boldsymbol{x} \\
\text{s.t.}\quad
& \boldsymbol{x}\in\Delta.
\end{aligned}
\end{equation}
The GMV model does not require the estimation of expected returns and depends
only on the covariance structure of asset returns. However, since variance
penalizes positive and negative deviations from the mean symmetrically, it
does not distinguish between upside and downside fluctuations.

\paragraph{Minimum-CVaR portfolio.}

CVaR focuses on the lower tail of the portfolio-return distribution. For a
tail probability level $\delta\in(0,1]$, we define
\begin{equation}
\label{eq:cvar-definition}
\operatorname{CVaR}_{\delta}\bigl(R_P(\boldsymbol{x})\bigr)
=
-\frac{1}{\delta}
\int_{0}^{\delta}Q_{R_P(\boldsymbol{x})}(\alpha)\,\mathrm{d}\alpha,
\end{equation}
where $Q_{R_P(\boldsymbol{x})}(\alpha)$ denotes the $\alpha$-quantile of the
portfolio return. Under this convention, CVaR is expressed as a loss measure,
so that larger values indicate greater downside risk.

In the discrete setting, let
\begin{equation*}
R_{P,(1)}(\boldsymbol{x})
\leq
R_{P,(2)}(\boldsymbol{x})
\leq
\cdots
\leq
R_{P,(T)}(\boldsymbol{x})
\end{equation*}
denote the ordered portfolio-return realizations. For each $j=1,\ldots,T$,
the empirical CVaR at the tail probability level $\delta_j=j/T$ is
\begin{equation}
\label{eq:empirical-cvar}
\operatorname{CVaR}_{\frac{j}{T}}
\bigl(R_P(\boldsymbol{x})\bigr)
=
-\frac{1}{j}
\sum_{t=1}^{j}R_{P,(t)}(\boldsymbol{x}).
\end{equation}
Thus, CVaR is the negative average of the $j$ smallest, and hence worst,
portfolio-return realizations.

Following \citet{rockafellar2000optimization}, the minimum-CVaR portfolio at
tail probability level $\delta$ is obtained by solving
\begin{equation}
\label{eq:minimum-cvar}
\begin{aligned}
\min_{\boldsymbol{x},\zeta,\boldsymbol{d}}\quad
& \zeta+\frac{1}{\delta T}\sum_{t=1}^{T}d_t \\
\text{s.t.}\quad
& d_t\geq-\sum_{k=1}^{n}r_{k,t}x_k-\zeta,
&& t=1,\ldots,T, \\
& d_t\geq0,
&& t=1,\ldots,T, \\
& \boldsymbol{x}\in\Delta, \\
& \zeta\in\mathbb{R}.
\end{aligned}
\end{equation}
The variable $\zeta$ represents a portfolio loss threshold, whereas $d_t$
measures the loss exceeding this threshold under scenario $t$. 
In our analysis, the minimum-CVaR portfolios are computed using a tail probability of
$\delta=0.05$.
Unlike the GMV
model, the minimum-CVaR model focuses directly on adverse portfolio outcomes.
Both models, however, minimize an aggregate measure of risk and may generate
portfolios concentrated in a relatively small number of assets.

\subsubsection{Risk-diversification strategies}\label{Sec:riskdiv}

Risk-diversification strategies do not focus exclusively on minimizing total
portfolio risk. Instead, they aim to distribute risk across the portfolio
components or to maximize the diversification benefit generated by combining
the assets. We consider the RP and MDP strategies, both based on portfolio
volatility.

Let
$
\sigma(\boldsymbol{x})
=
\sqrt{\boldsymbol{x}^{\top}\boldsymbol{\Sigma}\boldsymbol{x}}
$
denote portfolio volatility, and let
$\sigma_k=\sqrt{\sigma_{kk}}$ denote the volatility of asset $k$.

\paragraph{Risk Parity portfolio.}

Since volatility is differentiable and positively homogeneous of degree one,
Euler's theorem allows total portfolio volatility to be decomposed into the
risk contributions of the individual assets. The marginal risk contribution
of asset $k$ is
\begin{equation*}
\operatorname{MRC}_k(\boldsymbol{x})
=
\frac{\partial\sigma(\boldsymbol{x})}{\partial x_k}
=
\frac{(\boldsymbol{\Sigma}\boldsymbol{x})_k}
{\sigma(\boldsymbol{x})}.
\end{equation*}
The total risk contribution of asset $k$ is therefore
\begin{equation}
\label{eq:total-risk-contribution}
\operatorname{TRC}_k(\boldsymbol{x})
=
x_k\operatorname{MRC}_k(\boldsymbol{x})
=
\frac{x_k(\boldsymbol{\Sigma}\boldsymbol{x})_k}
{\sigma(\boldsymbol{x})}.
\end{equation}
The RP portfolio is characterized by equal total risk contributions:
\begin{equation}
\label{eq:risk-parity-condition}
\operatorname{TRC}_k(\boldsymbol{x}^{\mathrm{RP}})
=
\operatorname{TRC}_\ell(\boldsymbol{x}^{\mathrm{RP}}),
\qquad k,\ell=1,\ldots,n.
\end{equation}
Since the denominator in Eq.~\eqref{eq:total-risk-contribution} does not
depend on the asset index, Condition~\eqref{eq:risk-parity-condition} is
equivalent to
\begin{equation*}
x_k(\boldsymbol{\Sigma}\boldsymbol{x})_k
=
x_\ell(\boldsymbol{\Sigma}\boldsymbol{x})_\ell,
\qquad k,\ell=1,\ldots,n.
\end{equation*}

\noindent
Following \citet{maillard2010properties}, the RP portfolio can be computed by
solving the least-squares problem
\begin{equation}
\label{eq:risk-parity}
\begin{aligned}
\boldsymbol{x}^{\mathrm{RP}}
\in\arg\min_{\boldsymbol{x}}\quad
&
\sum_{k=1}^{n}\sum_{\ell=1}^{n}
\left[
x_k(\boldsymbol{\Sigma}\boldsymbol{x})_k
-x_\ell(\boldsymbol{\Sigma}\boldsymbol{x})_\ell
\right]^2 \\
\text{s.t.}\quad
& \boldsymbol{x}\in\Delta.
\end{aligned}
\end{equation}
If the optimal objective value in Problem~\eqref{eq:risk-parity} is zero,
all assets make the same contribution to portfolio volatility. More generally,
the objective function measures the aggregate discrepancy among the asset risk
contributions.

\paragraph{Most Diversified Portfolio.}

The MDP maximizes the Diversification Ratio introduced by
\citet{choueifaty2008towards}. For a long-only portfolio, this ratio is
defined as
\begin{equation}
\label{eq:diversification-ratio}
\operatorname{DR}(\boldsymbol{x})
=
\frac{\sum_{k=1}^{n}x_k\sigma_k}
{\sigma(\boldsymbol{x})}
=
\frac{\boldsymbol{x}^{\top}\boldsymbol{\sigma}}
{\sqrt{\boldsymbol{x}^{\top}\boldsymbol{\Sigma}\boldsymbol{x}}},
\end{equation}
where $\boldsymbol{\sigma}=(\sigma_1,\ldots,\sigma_n)^{\top}$. The numerator
is the weighted average of the individual asset volatilities, whereas the
denominator is the portfolio volatility, which accounts for the dependence
among asset returns. The Diversification Ratio therefore measures the relative
reduction in portfolio volatility obtained by combining the assets.

The MDP is obtained by solving
\begin{equation}
\label{eq:most-diversified-portfolio}
\begin{aligned}
\boldsymbol{x}^{\mathrm{MDP}}
\in\arg\max_{\boldsymbol{x}}\quad
& \operatorname{DR}(\boldsymbol{x}) \\
\text{s.t.}\quad
& \boldsymbol{x}\in\Delta.
\end{aligned}
\end{equation}
A larger Diversification Ratio indicates a greater diversification benefit.
Unlike RP, which equalizes the asset risk contributions, MDP directly
maximizes the relative reduction in portfolio volatility generated by the
combination of assets.

In summary, the four models represent two complementary portfolio selection
approaches. 
The GMV and minimum-CVaR portfolios minimize portfolio risk using variance and downside tail risk, respectively.
The RP portfolio equalizes the individual asset contributions to portfolio volatility, whereas the MDP maximizes the diversification benefit. 
These different
objectives generate distinct optimal weight structures. In the following
sections, we examine how these differences are reflected in the rank-size
representation of the portfolio weights.

\subsection{Rank-size analysis}\label{Sec:ranksize}

We use rank-size laws to describe the structure of the optimal portfolio
weights generated by the models introduced in Sect.~\ref{Sec:Portfolio_sel}.
A rank-size law relates the magnitude of an observation to its position in a
decreasingly ordered sequence. In our setting, it provides a parsimonious
representation of how portfolio capital is distributed from the largest to
the smallest allocation.

Let $\boldsymbol{x}=(x_1,\ldots,x_n)^{\top}\in\Delta$ be a vector of
portfolio weights. We denote its ordered components by
\begin{equation*}
x_{(1)}\geq x_{(2)}\geq\cdots\geq x_{(n)},
\end{equation*}
where $x_{(r)}$ is the weight associated with rank $r$. Thus, $r=1$
corresponds to the largest portfolio weight, whereas $r=n$ corresponds to the
smallest one.

For each optimal portfolio, the rank-size analysis consists of four steps:
\begin{enumerate}[noitemsep]
    \item sort the portfolio weights in decreasing order;
    \item fit each candidate rank-size function to the ordered weights;
    \item evaluate and compare the fitted functions using the criteria
    introduced in Sect.~\ref{Sec:evaluation};
    \item use the estimated parameters of the selected function to describe
    the portfolio weight structure and support the reverse-engineering
    analysis.
\end{enumerate}

We consider seven candidate specifications: the Exponential law (Exp), the
two-parameter Power law (Power), the shifted three-parameter Power law
(ShPower), the Discrete Generalized Beta Distribution (DGBD), the
Zipf--Mandelbrot law (ZML), the Polynomial law (Poly), and the Universal
law (Uni). Table~\ref{tab:rank-size_fun} reports their functional
forms, parameter restrictions, and main interpretations. To avoid confusion
with the asset-return variables, the maximum rank is denoted by $n$. Moreover,
the shift parameter of the Zipf--Mandelbrot law is denoted by $\eta$, since
$\varepsilon$ is used in Sect.~\ref{Sec:evaluation} as the tolerance parameter
of the CS$\varepsilon$SD criterion.

\begin{xltabular}{\textwidth}{
    >{\raggedright\arraybackslash}p{0.23\textwidth}
    >{\centering\arraybackslash}p{0.43\textwidth}
    >{\centering\arraybackslash}p{0.24\textwidth}}
\caption{Rank-size functions, parameters, and interpretations}
\label{tab:rank-size_fun}\\
\toprule
\textbf{Function} &
\textbf{Equation $y=f(r;\boldsymbol{\theta})$} &
\textbf{Parameter restrictions}\\
\midrule
\endfirsthead

\multicolumn{3}{c}{\tablename\ \thetable{} -- continued from previous page}\\
\toprule
\textbf{Function} &
\textbf{Equation $y=f(r;\boldsymbol{\theta})$} &
\textbf{Parameter restrictions}\\
\midrule
\endhead

\midrule
\multicolumn{3}{r}{Continued on next page}\\
\endfoot

\bottomrule
\endlastfoot

\textbf{Exponential law (Exp)}
&
$y=A\exp(-\alpha r)$
&
$A\geq0$, $\alpha\geq0$
\\
\multicolumn{3}{p{\dimexpr\textwidth-2\tabcolsep}}{
\textit{Interpretation.}
$A$ is a scale parameter, whereas $\alpha$ controls the rate at which the
weights decrease with rank. Larger values of $\alpha$ produce a steeper
profile \citep{Mandelbrot1997}.
}\\
\midrule

\textbf{Power law (Power)}
&
$\displaystyle y=\frac{B}{r^{\beta}}$
&
$B\geq0$, $\beta\geq0$
\\
\multicolumn{3}{p{\dimexpr\textwidth-2\tabcolsep}}{
\textit{Interpretation.}
$B$ determines the scale and coincides with the fitted value at the first
rank. The exponent $\beta$ controls the rate of decay and the weight assigned
to the tail of the ranked distribution \citep{pareto1896cours}.
}\\
\midrule

\textbf{Shifted Power law (ShPower)}
&
$\displaystyle y=\frac{B}{r^{\beta}}+C$
&
$B\geq0$, $\beta\geq0$, $C\in\mathbb{R}$
\\
\multicolumn{3}{p{\dimexpr\textwidth-2\tabcolsep}}{
\textit{Interpretation.}
This specification extends the Power law by adding the vertical-shift
parameter $C$. It can therefore accommodate ranked profiles that approach a
nonzero level over the observed rank domain.
}\\
\midrule

\textbf{Discrete Generalized Beta Distribution (DGBD)}
&
$\displaystyle y=D\frac{(n+1-r)^{\gamma}}{r^{\xi}}$
&
$D\geq0$, $\gamma\geq0$, $\xi\geq0$
\\
\multicolumn{3}{p{\dimexpr\textwidth-2\tabcolsep}}{
\textit{Interpretation.}
$D$ is a scale parameter. The parameters $\gamma$ and $\xi$ control the
curvature at the lower- and upper-rank portions of the profile, respectively.
When $\gamma=0$, the specification reduces to a Power law
\citep{naumis2008tail,martinez2009universality}.
}\\
\midrule

\textbf{Zipf--Mandelbrot law (ZML)}
&
$\displaystyle y=\frac{E}{(r+\eta)^{\lambda}}$
&
$E\geq0$, $\eta\geq0$, $\lambda\geq0$
\\
\multicolumn{3}{p{\dimexpr\textwidth-2\tabcolsep}}{
\textit{Interpretation.}
$E$ determines the scale, $\eta$ regulates the behavior at the first ranks,
and $\lambda$ controls the rate of decay. When $\eta=0$, the specification
reduces to the standard Zipf form \citep{mandelbrot1953informational,
mandelbrot1961theory}.
}\\
\midrule

\textbf{Polynomial law (Poly)}
&
$y=p_1r^3+p_2r^2+p_3r+p_4$
&
$p_i\in\mathbb{R}$, $i=1,\ldots,4$
\\
\multicolumn{3}{p{\dimexpr\textwidth-2\tabcolsep}}{
\textit{Interpretation.}
The four coefficients provide a flexible cubic approximation of the ranked
weights. Unlike the other specifications, the Polynomial law is not based on
a scaling relation and its coefficients do not have a direct tail-decay
interpretation \citep{CERQUETI2022111975}.
}\\
\midrule

\textbf{Universal law (Uni)}
&
$\displaystyle
y=F\frac{(n+1-r+\Psi)^{\zeta_1}}
{[n(r+\Phi)]^{\zeta_2}}$
&
$F,\Psi,\Phi,\zeta_1,\zeta_2\geq0$
\\
\multicolumn{3}{p{\dimexpr\textwidth-2\tabcolsep}}{
\textit{Interpretation.}
$F$ is a scale parameter. The exponents $\zeta_1$ and $\zeta_2$ determine
the curvature of the ranked profile, whereas $\Phi$ and $\Psi$ regulate its
behavior at the smallest and largest ranks, respectively
\citep{ausloos2016universal}.
}\\

\end{xltabular}

For each candidate function $f$, let $\boldsymbol{\theta}_f$ denote its
parameter vector and let $\Theta_f$ denote the corresponding feasible
parameter space reported in Table~\ref{tab:rank-size_fun}. We estimate the
parameters by solving the nonlinear least-squares problem
\begin{equation}
\label{eq:least-squaresProblem}
\widehat{\boldsymbol{\theta}}_f
=
\underset{\boldsymbol{\theta}_f\in\Theta_f}{\operatorname{arg\,min}}
\sum_{r=1}^{n}
\left[
x_{(r)}-f\!\left(r;\boldsymbol{\theta}_f\right)
\right]^2.
\end{equation}
The fitted value associated with rank $r$ is therefore
\begin{equation*}
\widehat{x}_{(r)}^{(f)}
=
f\!\left(r;\widehat{\boldsymbol{\theta}}_f\right).
\end{equation*}

We solve Problem~\eqref{eq:least-squaresProblem} using the Trust Region
Reflective (TRR) algorithm \citep{branch1999subspace}. The parameter bounds
are imposed according to the restrictions reported in
Table~\ref{tab:rank-size_fun}. In particular, the shift parameter $C$ of the
shifted Power law and the coefficients of the Polynomial law are allowed to
take real values.

To improve numerical convergence and reduce sensitivity to the initial point, we first perform a grid search over the feasible parameter space $\Theta_f$.
Among the parameter vectors evaluated on the grid, we select the one producing
the lowest RMSE and use it to initialize the TRR algorithm. This procedure is
repeated for every candidate function, portfolio model, and rolling estimation
window. The resulting goodness-of-fit values are compared using the framework
introduced in Sect.~\ref{Sec:evaluation}, whereas the estimated parameters of
the selected rank-size function are used in the reverse-engineering analysis.


\subsection{Goodness-of-fit evaluation framework}\label{Sec:evaluation}

The rank-size literature does not provide a universally accepted criterion
for selecting among competing functional forms. Moreover, relying on a single
goodness-of-fit measure may provide only a partial assessment. We therefore
evaluate each fitted function using four complementary measures: the \textit{degrees-of-freedom-adjusted} Root Mean
Squared Error (RMSE), the Mean Absolute Error (MAE), the coefficient of
determination ($R^2$), and the Akaike Information Criterion (AIC). RMSE and
MAE measure the distance between the observed and fitted weights, $R^2$
measures the proportion of observed variability explained by the fitted
function, and AIC balances fitting accuracy and model complexity.

\paragraph{Goodness-of-fit measures.}

Let $x_{(r)}$ denote the observed portfolio weight associated with rank $r$,
and let
$
\widehat{x}_{(r)}^{(f)}
=
f\!\left(r;\widehat{\boldsymbol{\theta}}_f\right)
$
denote the corresponding value predicted by rank-size function $f$, for
$r=1,\ldots,n$. For ease of notation, we omit the superscript $(f)$ when no
confusion can arise. 

\noindent
The goodness-of-fit measures are defined as
\begin{align}
\operatorname{RMSE}
&=
\sqrt{
\frac{1}{n-p_f}
\sum_{r=1}^{n}
\left(x_{(r)}-\widehat{x}_{(r)}\right)^2
},
\label{eq:rmse_definition}
\\
\operatorname{MAE}
&=
\frac{1}{n}
\sum_{r=1}^{n}
\left|x_{(r)}-\widehat{x}_{(r)}\right|,
\label{eq:mae_definition}
\\
R^2
&=
1-
\frac{
\sum_{r=1}^{n}
\left(x_{(r)}-\widehat{x}_{(r)}\right)^2
}{
\sum_{r=1}^{n}
\left(x_{(r)}-\overline{x}\right)^2
},
\label{eq:r2_definition}
\end{align}
where
$
\overline{x}
=
\frac{1}{n}\sum_{r=1}^{n}x_{(r)}
$
and $p_f$ is the number of parameters estimated for function $f$.

The RMSE in Eq.~\eqref{eq:rmse_definition} is adjusted for the number of estimated parameters by using $n-p_f$ instead of $n$ in the denominator. This adjustment accounts for differences in model complexity and mitigates the tendency of more flexible specifications to achieve lower in-sample fitting errors. As a robustness check, using the conventional denominator $n$ does not change the rank-size law selected by the CS$\varepsilon$SD criterion in any of the four datasets.
Lower RMSE and MAE values indicate a more accurate fit, whereas a larger
$R^2$ indicates greater explanatory power.

\noindent
Assuming independent Gaussian fitting errors with constant variance, AIC is
computed, up to an additive constant common to all candidate functions, as
\begin{equation}
\label{eq:aic_definition}
\operatorname{AIC}
=
n\log\!\left(\frac{\operatorname{RSS}}{n}\right)+2p_f,
\qquad
\operatorname{RSS}
=
\sum_{r=1}^{n}
\left(x_{(r)}-\widehat{x}_{(r)}\right)^2,
\end{equation}
where $\operatorname{RSS}$ is the residual sum of squares.
A lower AIC indicates a better trade-off between fitting accuracy and model complexity.

Each measure is computed separately for every candidate rank-size function,
portfolio model, and rolling estimation window. Therefore, for a given
rank-size function, the empirical procedure described in
Sect.~\ref{Sec:empanalysis} produces a distribution of goodness-of-fit values
rather than a single value. A conventional comparison based on means,
standard deviations, medians, minima, and maxima may yield conflicting
rankings. For example, one function may have a lower average error, whereas
another may have lower variability or better worst-case performance. In such
cases, selecting the best function from summary statistics alone becomes
ambiguous.

To address this issue, we compare the entire empirical distributions of the
fitting errors using stochastic dominance. The methodology is presented for
RMSE, but the same procedure can be applied to MAE or any other error measure
for which smaller values are preferred.

\paragraph{Stochastic dominance.}

Let $A$ and $B$ be two random variables with cumulative distribution
functions
$
F_A(\alpha)=\mathbb{P}(A\leq\alpha)$ and $F_B(\alpha)=\mathbb{P}(B\leq\alpha)$, with $\alpha\in\mathbb{R}$.

\begin{definition}[First-order Stochastic Dominance]
\label{Def:FSD}
The random variable $A$ weakly First-order Stochastic Dominates $B$, denoted
by $A\succeq_{\mathrm{FSD}}B$, if
\begin{equation*}
F_A(\alpha)\leq F_B(\alpha),
\qquad \forall\alpha\in\mathbb{R}.
\end{equation*}
The dominance relation is strict if the inequality is strict for at least one
$\alpha$.
\end{definition}

\begin{definition}[Second-order Stochastic Dominance]
\label{Def:SSD}
The random variable $A$ weakly Second-order Stochastic Dominates $B$, denoted
by $A\succeq_{\mathrm{SSD}}B$, if
\begin{equation*}
\int_{-\infty}^{\alpha}F_A(\tau)\,\mathrm{d}\tau
\leq
\int_{-\infty}^{\alpha}F_B(\tau)\,\mathrm{d}\tau,
\qquad \forall\alpha\in\mathbb{R}.
\end{equation*}
The dominance relation is strict if the inequality is strict for at least one
$\alpha$.
\end{definition}

\noindent
These relations have a direct Expected Utility Theory interpretation
\citep{levy1992stochastic}. In particular, $A\succeq_{\mathrm{FSD}}B$ if and
only if
$
\mathbb{E}[u(A)]\geq\mathbb{E}[u(B)]
$
for every non-decreasing utility function $u$. Similarly,
$A\succeq_{\mathrm{SSD}}B$ if and only if the same inequality holds for every
non-decreasing and concave utility function. FSD implies SSD, whereas the
converse does not generally hold \citep[see, e.g.,][]{levy2015stochastic}. In
this paper, we focus on SSD and its cumulative relaxation
\citep[see, e.g.,][]{lizyayev2012tractable,cesarone2025flexible}.

\paragraph{Application to empirical RMSE distributions.}

Let $Q$ denote the number of rolling estimation windows, and hence the number
of portfolio rebalancing dates, and let $M$ denote the number of portfolio
selection models. For each candidate rank-size function $f$, portfolio model
$m=1,\ldots,M$, and rebalancing date $q=1,\ldots,Q$, we obtain one RMSE value
from the fit of the corresponding ranked optimal portfolio weights. When the
errors are pooled across portfolio models, the empirical RMSE distribution of
each rank-size function contains
$
N=MQ
$
observations. When the comparison is conducted separately for a given
portfolio model, $N=Q$.

Let $f$ and $g$ be two candidate rank-size functions. Since smaller RMSE
values indicate better fitting performance, we define
\begin{equation}
\label{eq:performance_variables}
X_i=-\operatorname{RMSE}_i^{(f)}
\qquad\text{and}\qquad
Y_i=-\operatorname{RMSE}_i^{(g)},
\qquad i=1,\ldots,N.
\end{equation}
Thus, $X$ and $Y$ are fitting-performance variables for which larger
realizations are preferred.

Following \citet{cesarone2025flexible}, let
$
X_{(1)}\leq X_{(2)}\leq\cdots\leq X_{(N)}$ and $Y_{(1)}\leq Y_{(2)}\leq\cdots\leq Y_{(N)}
$
denote the realizations of $X$ and $Y$, respectively, sorted in increasing
order. Since $X_i$ and $Y_i$ are the negatives of the corresponding fitting
errors, their smallest realizations correspond to the largest, and hence
worst, RMSE values.

\noindent
We assign equal probability to the $N$ empirical realizations:
$
\mathbb{P}(X=X_i)
=
\mathbb{P}(Y=Y_i)
=
\frac{1}{N}$,
with $i=1,\ldots,N$.
The same construction can be extended to a general discrete probability
distribution, as shown by \citet{cesarone2025flexible}.

\noindent
Under equal probabilities, the stochastic dominance conditions simplify as
follows:
\begin{itemize}[noitemsep]
    \item $X\succeq_{\mathrm{FSD}}Y$ if and only if
    $
    X_{(j)}\geq Y_{(j)}$,
    $\forall j=1,\ldots,N$;

    \item $X\succeq_{\mathrm{SSD}}Y$ if and only if
    $
    \sum_{i=1}^{j}X_{(i)}
    \geq
    \sum_{i=1}^{j}Y_{(i)}$,
    $\forall j=1,\ldots,N$.
\end{itemize}
In each case, the dominance relation is strict when at least one of the
corresponding inequalities is strict.

\noindent
For $j=1,\ldots,N$, define the empirical lower-tail functional
$
\operatorname{Tail}_{\frac{j}{N}}(X)
=
\frac{1}{N}\sum_{i=1}^{j}X_{(i)}.
$
Using the CVaR convention adopted by \citet{cesarone2025flexible}, we also
have
$
\operatorname{CVaR}_{\frac{j}{N}}(X)
=
-\frac{1}{j}\sum_{i=1}^{j}X_{(i)}.
$
Consequently,
\begin{equation}
\label{eq:tail_cvar_relation}
\operatorname{Tail}_{\frac{j}{N}}(X)
=
-\frac{j}{N}
\operatorname{CVaR}_{\frac{j}{N}}(X).
\end{equation}
The negative sign follows from the CVaR convention for return variables,
under which CVaR is the negative average of the lower-tail realizations. The
same relations hold for $Y$.

\noindent
Since $X_i=-\operatorname{RMSE}_i^{(f)}$,
$\operatorname{CVaR}_{j/N}(X)$ is the average of the $j$ largest RMSE values
generated by function $f$. It therefore measures the average fitting error
among the $j$ worst realizations. The analogous interpretation holds for
$\operatorname{CVaR}_{j/N}(Y)$ and function $g$.

\noindent
It follows from Eq.~\eqref{eq:tail_cvar_relation} that
$X\succeq_{\mathrm{SSD}}Y$ if and only if
\begin{equation}
\label{eq:SSD_CVaR}
\operatorname{CVaR}_{\frac{j}{N}}(X)
\leq
\operatorname{CVaR}_{\frac{j}{N}}(Y),
\qquad j=1,\ldots,N.
\end{equation}
For strict SSD, at least one inequality in Eq.~\eqref{eq:SSD_CVaR} must be
strict. Thus, function $f$ is preferred to function $g$ under SSD when, for
every tail size $j$, its average over the $j$ worst RMSE values is no larger
than the corresponding average for function $g$.

\paragraph{Cumulative Second-order $\varepsilon$-Stochastic Dominance.}

Exact SSD may fail when the empirical CVaR profiles cross, even if the
violations are small. Approximate stochastic dominance relations allow these violations to be controlled rather than requiring all the exact inequalities
\citep[see, e.g.,][]{leshno2002preferred}. In this setting, we use the Cumulative Second-order $\varepsilon$-Stochastic
Dominance (CS$\varepsilon$SD) criterion introduced by \citet{cesarone2025flexible}. 
It is a cumulative refinement of approximate SSD and is methodologically related to the
Cumulative Zero-order $\varepsilon$-Stochastic Dominance relation of
\citet{bruni2017exact}.

\begin{definition}[Cumulative Second-order $\varepsilon$-Stochastic Dominance]
\label{def:CSepsilonSD}
Let $\varepsilon\geq0$ and $\mathcal{J}=\{1,\ldots,N\}$. The
fitting-performance variable $X$ is preferred to $Y$ with respect to
CS$\varepsilon$SD if
\begin{equation}
\label{eq:CSepsilonSD_definition}
\sum_{k\in S}
\left[
\operatorname{CVaR}_{\frac{k}{N}}(X)
-
\operatorname{CVaR}_{\frac{k}{N}}(Y)
\right]
\leq\varepsilon,
\qquad \forall S\subseteq\mathcal{J}.
\end{equation}
\end{definition}

\noindent
For $\varepsilon=0$, Definition~\ref{def:CSepsilonSD} reduces to weak SSD,
because it includes all singleton subsets $S=\{k\}$. Positive values of
$\varepsilon$ allow controlled cumulative violations of the CVaR inequalities
in Eq.~\eqref{eq:SSD_CVaR}.

\noindent
Directly checking Definition~\ref{def:CSepsilonSD} would require an
exponential number of inequalities. However, \citet[Proposition~1]
{cesarone2025flexible} provide an equivalent polynomial-time condition. 

\noindent
For
each $k\in\mathcal{J}$, define
\begin{equation*}
d_k(X,Y)
=
\operatorname{CVaR}_{\frac{k}{N}}(X)
-
\operatorname{CVaR}_{\frac{k}{N}}(Y).
\end{equation*}
Let
\begin{equation*}
d^{(1)}(X,Y)
\geq
d^{(2)}(X,Y)
\geq
\cdots
\geq
d^{(N)}(X,Y)
\end{equation*}
denote these differences sorted in non-increasing order. The superscript in
parentheses distinguishes the ordered CVaR differences from the ordered
realizations $X_{(i)}$ and $Y_{(i)}$. Then, $X$ is preferred to $Y$ with
respect to CS$\varepsilon$SD if and only if
\begin{equation}
\label{eq:CS_epsilon_SD}
\sum_{j=1}^{t}d^{(j)}(X,Y)
\leq\varepsilon,
\qquad t=1,\ldots,N.
\end{equation}
Indeed, among all subsets of cardinality $t$, the sum of the $t$ largest CVaR
differences produces the largest cumulative violation.

\noindent
Rather than fixing $\varepsilon$ ex-ante, we define the signed dominance
index
\begin{equation}
\label{eq:signed_dominance_index}
\mathcal{D}(X,Y)
=
\max_{t=1,\ldots,N}
\sum_{j=1}^{t}d^{(j)}(X,Y).
\end{equation}
The index $\mathcal{D}(X,Y)$ is the largest cumulative CVaR difference in
Eq.~\eqref{eq:CS_epsilon_SD}. Hence,
\begin{equation*}
\mathcal{D}(X,Y)\leq0
\quad\Longleftrightarrow\quad
X\succeq_{\mathrm{SSD}}Y.
\end{equation*}
If $\mathcal{D}(X,Y)<0$, all the CVaR inequalities are strict. If
$\mathcal{D}(X,Y)=0$, weak SSD holds; the SSD relation is strict whenever at
least one inequality in Eq.~\eqref{eq:SSD_CVaR} is strict.

\noindent
When $\mathcal{D}(X,Y)>0$, exact SSD does not hold, but $X$
CS$\varepsilon$SD-dominates $Y$ for every
$\varepsilon\geq\mathcal{D}(X,Y)$. Therefore, the minimum non-negative
tolerance is
\begin{equation}
\label{eq:minimum_tolerance}
\varepsilon_{\min}(X,Y)
=
\max\left\{0,\mathcal{D}(X,Y)\right\}.
\end{equation}

\noindent
In the empirical analysis, the CVaR differences and the signed dominance index are computed on the full empirical grid of tail probability levels \(\delta_k=k/N\), \(k=1,\ldots,N\), where \(CVaR_{k/N}(X)\) is the average of the \(k\) largest RMSE values. 
Hence, all the inequalities in Eq. \eqref{eq:SSD_CVaR} are checked, and \(D(X,Y)\leq 0\) is equivalent to exact empirical SSD.

\noindent
For each pair of candidate functions $f$ and $g$, we compute the signed
dominance indices in both directions, $\mathcal{D}(X,Y)$ and
$\mathcal{D}(Y,X)$. Function $f$ is preferred to function $g$ when
\begin{equation}
\label{eq:pairwise_preference}
\mathcal{D}(X,Y)
<
\mathcal{D}(Y,X).
\end{equation}
The reverse inequality favors function $g$, whereas equality indicates that
the pairwise comparison does not distinguish between them. A smaller index
indicates stronger dominance when exact SSD holds, or a smaller cumulative
violation when it does not.

\noindent
We repeat this comparison for all pairs of candidate rank-size functions. If
a function is preferred, according to Eq.~\eqref{eq:pairwise_preference}, to
every competing function, we select it as the pairwise winner. Its calibrated
parameters are then used as input features in the reverse-engineering analysis
described in Sect.~\ref{Sec:reverse}, where the selection is repeated within
each training set.

\subsection{Reverse engineering via Random Forest classification}\label{Sec:reverse}

The portfolio models introduced in Sect.~\ref{Sec:Portfolio_sel} pursue
different objectives and may therefore generate different rank-size profiles.
We now consider the inverse problem: given the parameters of the rank-size
function fitted to a vector of optimal weights, we assess whether they
identify the portfolio model that generated the weights.

\noindent
This problem arises, for example, when an analyst observes the disclosed
weights of a portfolio but not the rule used to construct it. The analyst can
fit the selected rank-size function to the sorted weights and use the
estimated parameters as a compact description of the portfolio geometry. A
classifier trained on portfolios with known generating models can then be
used to infer the construction rule.

\paragraph{Classification data.}

Let $f^{\star}$ denote the rank-size function selected by the procedure
described in Sect.~\ref{Sec:evaluation}. For each portfolio model
$m=1,\ldots,M$ and rolling window $q=1,\ldots,Q$, let
$
\boldsymbol{z}_{m,q}
=
\widehat{\boldsymbol{\theta}}_{f^{\star},m,q}
\in\mathbb{R}^{p_{f^{\star}}}
$
denote the parameters obtained by fitting $f^{\star}$ to the sorted optimal
weights of model $m$ in window $q$. The vector $\boldsymbol{z}_{m,q}$ is the
feature vector of observation $(m,q)$, and the dataset contains $N=MQ$
observations.

\noindent
We consider two classification tasks and denote by $c_{m,q}$ the label of
observation $(m,q)$ and by $\mathcal{C}$ the set of possible labels. In the
model-level task, the label is the generating model and
$\mathcal{C}=\{\mathrm{GMV},\mathrm{CVaR},\mathrm{RP},\mathrm{MDP}\}$. In the
strategy-level task, the label is the class of the generating model and
$\mathcal{C}=\{\text{minimum-risk},\ \text{risk-diversification}\}$, where GMV
and CVaR belong to the first class and RP and MDP to the second. A separate
classifier is trained for each task.

\paragraph{Random Forest classifier.}

We use a Random Forest (RF) classifier \citep{breiman2001random}. The RF
consists of $N_{\mathrm{tree}}$ classification trees, each grown on a
bootstrap sample of the training observations and using, at each split, a
randomly selected subset of the features. For a feature vector
$\boldsymbol{z}$, let $\widehat{\pi}_b(c\mid\boldsymbol{z})$ denote the
proportion of training observations with label $c$ in the leaf of tree $b$
that contains $\boldsymbol{z}$. The RF prediction is the label with the
highest average proportion across trees:
\begin{equation}
\label{eq:rf_vote}
\widehat{c}(\boldsymbol{z})
=
\underset{c\in\mathcal{C}}{\operatorname{arg\,max}}\,
\frac{1}{N_{\mathrm{tree}}}
\sum_{b=1}^{N_{\mathrm{tree}}}
\widehat{\pi}_b(c\mid\boldsymbol{z}).
\end{equation}
Averaging over bootstrap samples reduces the variance of individual
trees, while each tree can capture nonlinear relations between the parameters
and the labels.

\paragraph{Validation design.}

The observations are not independent. Consecutive estimation windows share
most of their daily returns, so the parameters fitted in window $q$ are close
to those fitted in neighbouring windows. Moreover, the $M$ portfolios of the
same window are computed from the same returns. If individual observations
were randomly assigned to training and test sets, the test set would contain
portfolios almost identical to training portfolios, and the resulting
accuracy would overstate the ability of the classifier to generalize to new
periods.

\noindent
We therefore assign entire windows to the training and test sets and separate
them by an embargo. Let $L$ denote the number of daily returns in each
estimation window and $s$ the number of trading days by which the window is
shifted. Windows $q$ and $q'$ share $\max\{L-s|q-q'|,0\}$ daily returns, so
the embargo
\begin{equation}
\label{eq:embargo}
h=\left\lceil\frac{L}{s}\right\rceil-1
\end{equation}
is the largest distance between two windows that share at least one return.

\noindent
We evaluate the classifier by blocked cross-validation. The windows
$1,\ldots,Q$ are partitioned into $V$ blocks of consecutive windows. For
block $v$, with first window $\underline{q}_v$ and last window
$\overline{q}_v$, the test set contains the $M$ portfolios of each window in
the block, whereas the training set contains the portfolios of the windows
$q\leq\underline{q}_v-h-1$ and $q\geq\overline{q}_v+h+1$. Each observation is
therefore tested exactly once. Within each training set, the selection
procedure of Sect.~\ref{Sec:evaluation} is repeated using only the training
windows, so that the choice of $f^{\star}$ does not depend on the test
observations. The RF settings are fixed in advance and are not tuned on the
test sets. As robustness checks, we also consider a walk-forward validation,
which trains only on windows preceding the test block, with the same embargo,
and a random holdout that assigns individual observations to the training and
test sets irrespective of their window.

\paragraph{Performance measures.}
In each test block, we compute the accuracy, defined as the fraction of
correctly classified observations, and the recall of each class
$c\in\mathcal{C}$, defined as the fraction of observations with label $c$
that are correctly classified. We report the mean and the standard deviation
of the accuracy across the $V$ blocks, together with an approximate $95\%$
confidence interval based on the Student $t$ distribution with $V-1$ degrees
of freedom. Since the training sets of different blocks overlap, the block
accuracies are not independent, and the interval should be interpreted as
indicative \citep{bengio2004no}. We also report the confusion matrix obtained
by pooling the predictions of all test blocks.

\section{Empirical analysis}\label{Sec:empanalysis}

The experiments have been conducted on real-world datasets downloaded from
\emph{LSEG Data $\&$ Analytics} \citep[formerly known as
\emph{Refinitiv-Datastream}, see][]{LSEG}.
The data consists of daily prices, adjusted for dividends and stock splits, of the constituents of four major stock indices, namely the Dow Jones Industrial Average (DJI), the EURO STOXX 50 (STOXX50E), the NASDAQ 100 (NDX), and the FTSE 250, listed in Table~\ref{tab:datasets} and publicly available at \url{https://www.francescocesarone.com/data-sets}.
For each index, we consider the constituents as of December 2023, the end of the sample period, and retain only those with a complete price history from
January 2009 to December 2023. Prices are observed on all weekdays, and on market holidays the last available price is carried forward, so that each series contains 3912 daily observations. After these filters, the final samples contain 29 assets for the DJI, 47 for the STOXX50E, 72 for the NDX, and 166 for the FTSE 250.
\begin{table}[htbp!]
    \centering
    \begin{tabular}{c c c c c}
    \hline
    Index & \# Assets & \# Days & Time interval & Abbreviation\\
    \hline
     Dow Jones & 29 & 3912 & Jan 2009-Dec 2023 & DJI\\
    \hline
     EURO STOXX 50 & 47 & 3912 & Jan 2009-Dec 2023 & STOXX50E\\
     \hline
    NASDAQ 100 & 72 & 3912 & Jan 2009-Dec 2023 & NDX\\
     \hline
     FTSE 250 & 166 & 3912 & Jan 2009-Dec 2023 & FTSE 250\\
     \hline
    \end{tabular}
    \caption{Equity datasets used in the empirical analysis. Prices are adjusted
for dividends and stock splits, and the sample covers the period from
January 2009 to December 2023.}
    \label{tab:datasets}
\end{table}

\noindent
We present the detailed analysis for the FTSE 250 dataset, while results for the remaining markets are reported in Appendix \ref{Sec: additional_results}.

\noindent
We use a rolling-window design. Each estimation window contains \(L=250\) daily returns and is shifted forward by \(s=20\) trading days. We retain only estimation windows for which at least one subsequent interval of \(s\) daily returns remains available within the sample. Given the 3911 daily returns, this convention yields \(Q=183\) windows, with 21 daily returns remaining after the final estimation window.

In each window, we compute the GMV, minimum-CVaR, RP, and MDP
portfolios. We then fit the seven candidate rank-size functions to the sorted
portfolio weights and record $R^2$, RMSE, MAE, and AIC.

For each candidate function, we pool these measures across the $M=4$
portfolio models and the $Q$ rolling windows, obtaining $N=MQ=732$
observations for each measure. This pooled analysis provides an overall comparison across portfolio
models. 

\begin{figure}
    \centering
    \renewcommand{\thesubfigure}{}

    \subfigure[ZML fit to GMV weights]{
        \includegraphics[width=0.45\textwidth]
        {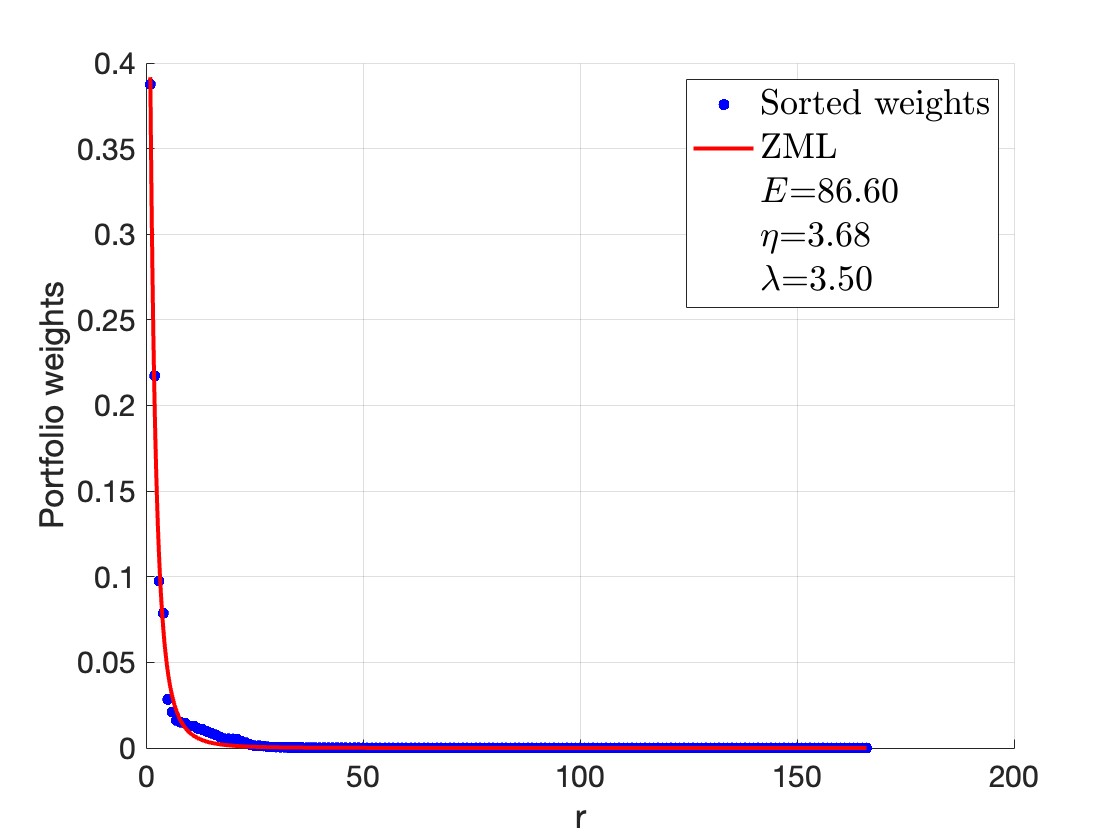}
    }
    \subfigure[ZML fit to minimum-CVaR weights]{
        \includegraphics[width=0.45\textwidth]
        {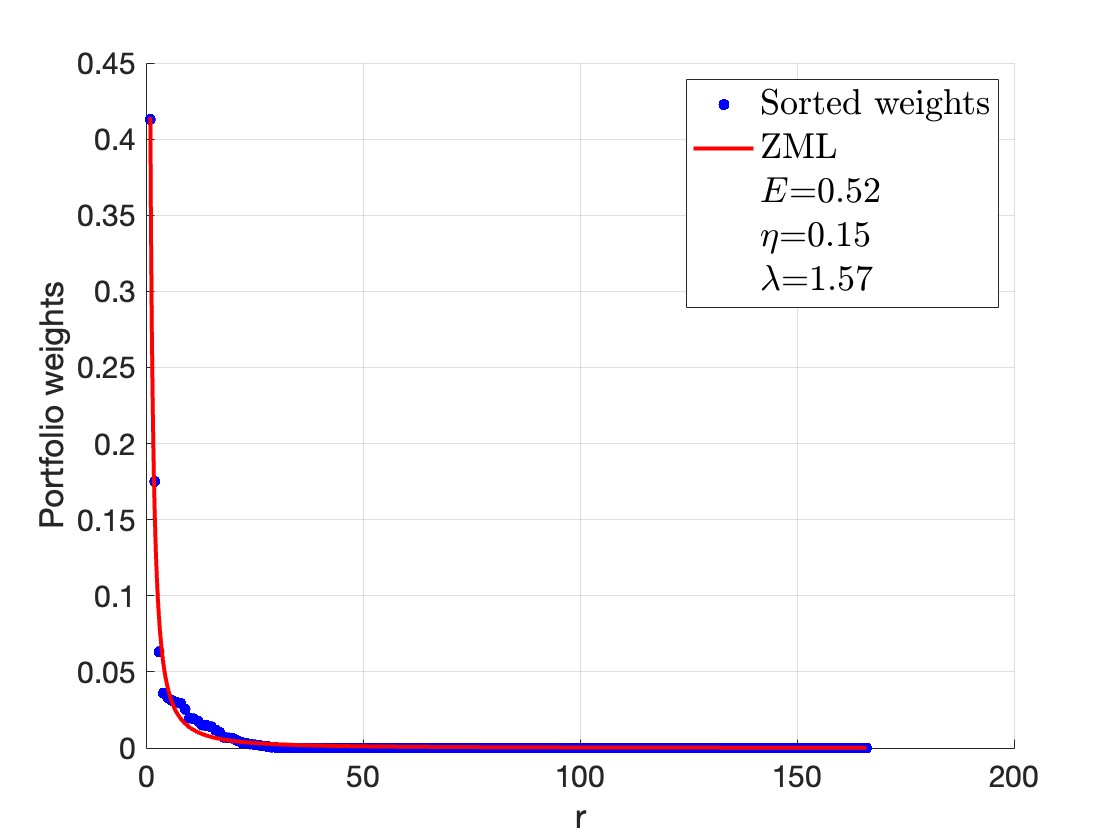}
    }

    \subfigure[ZML fit to RP weights]{
        \includegraphics[width=0.45\textwidth]
        {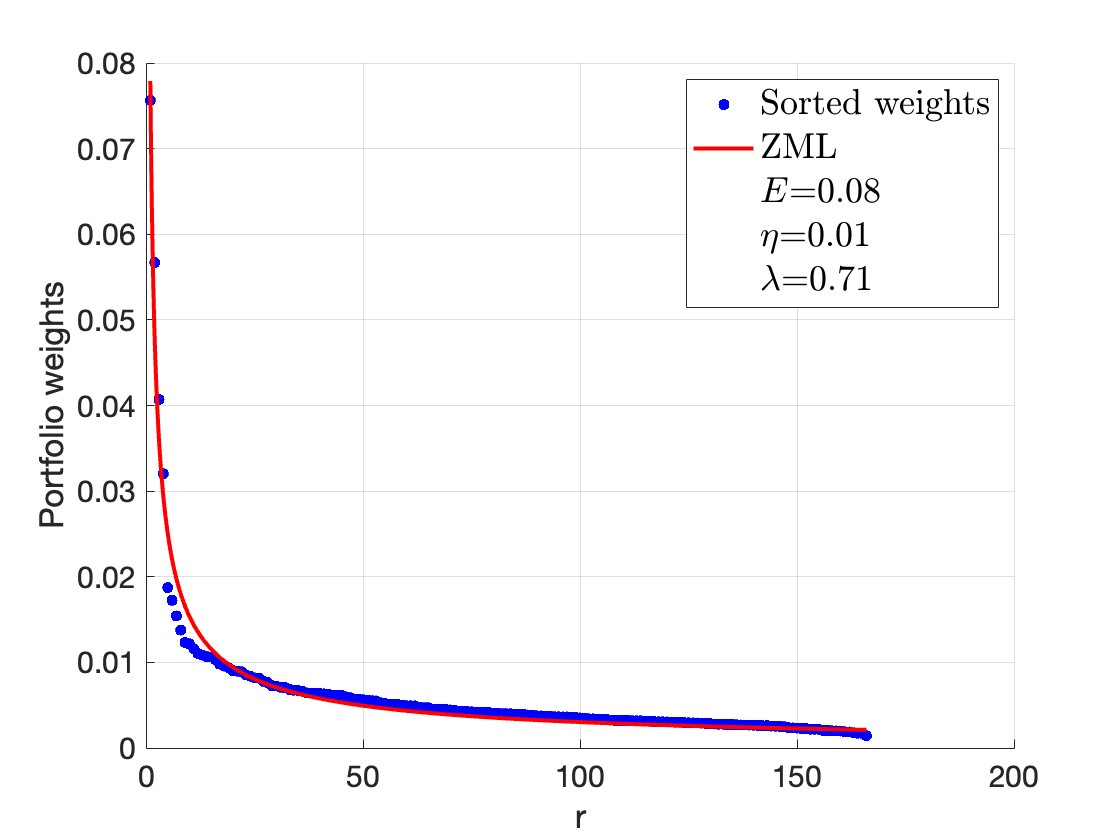}
    }
    \subfigure[ZML fit to MDP weights]{
        \includegraphics[width=0.45\textwidth]
        {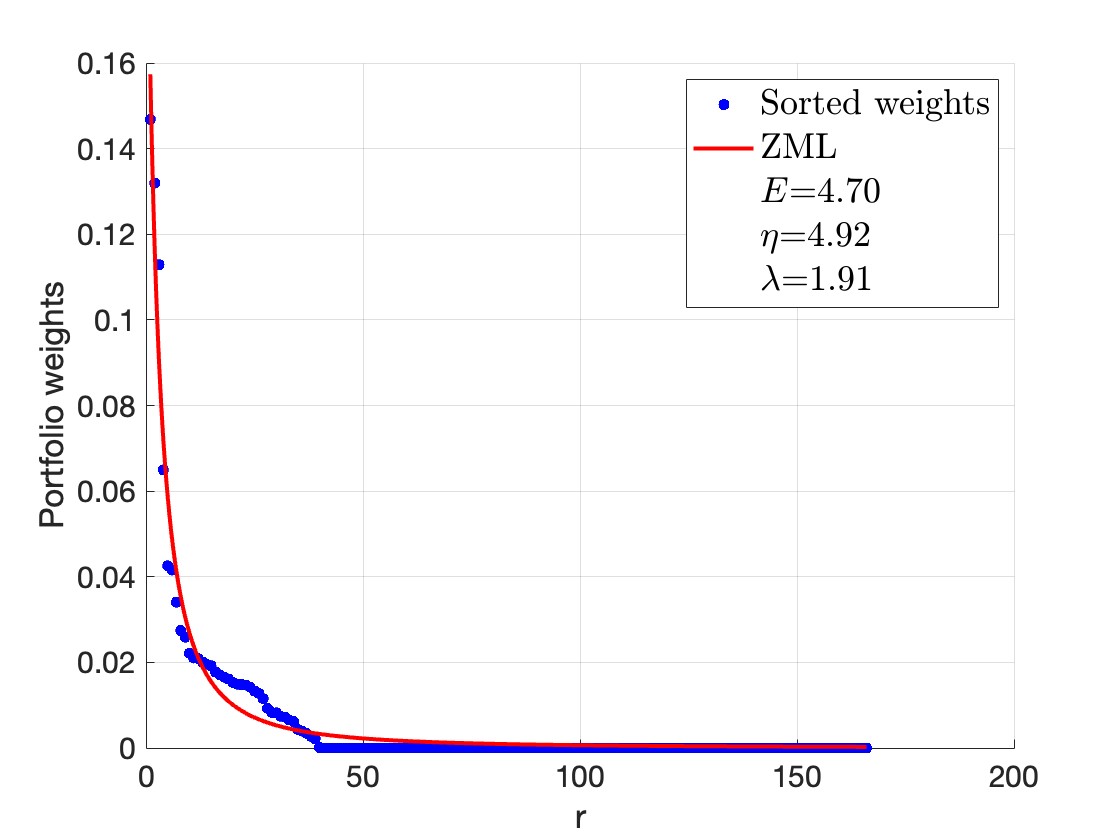}
    }

    \caption{ZML fits to the sorted portfolio weights obtained from the GMV, minimum-CVaR, RP, and MDP models for one rolling window
    of the FTSE 250 dataset.}
    \label{fig:zml-fits-ftse250}
\end{figure}

\noindent
For illustrative purposes, Figure~\ref{fig:zml-fits-ftse250} shows the optimal portfolio weights sorted in descending order and the fitted ZML function for a single in-sample window from the FTSE 250 dataset, across all four portfolio selection models.
This fitting procedure is repeated for each in-sample window, for each model, and for all candidate rank-size functions.
%

\subsection{Comparative results}

In this section, we analyze the performance of the seven rank-size functions considered in fitting the optimal portfolio weights generated for the FTSE dataset.
To provide a robust assessment across the $N=732$ portfolio weight distributions generated by the rolling window procedure, we rely on standard summary statistics of the goodness-of-fit measures: mean ($\mu$), standard deviation ($\sigma$), median ($Me$), minimum ($\min$), and maximum ($\max$).
The ranking of the summary statistics is shown through different colors. More precisely, for each column, the colors range from deep green to deep red, where the former represents the best value and the latter indicates the worst one.
The mean values reported in Tables~\ref{tab:R2_FT250},
\ref{tab:RMSE_FT250}, \ref{tab:MAE_FT250}, and \ref{tab:AIC_FT250} rank the
ZML first among the candidate functions according to all four
criteria in the FTSE 250 sample.

\noindent
Regarding explanatory power, Table~\ref{tab:R2_FT250} shows that the
ZML has the highest mean $R^2$ ($\mu\approx0.97$) and the
lowest standard deviation ($\sigma\approx0.027$) among the candidate
functions.
Its median, minimum, and maximum $R^2$ values are approximately $0.98$,
$0.78$, and $0.9996$, respectively.
The ShPower, DGBD, and Uni have mean $R^2$ values
exceeding $0.95$. Among the candidate functions, the Poly has the
lowest mean $R^2$ ($\mu\approx0.58$) and the highest standard deviation
($\sigma\approx0.22$). Its minimum $R^2$ is approximately $0.155$, indicating
a comparatively weak fit for some portfolio instances.

\noindent
The error-metric analysis, presented in Tables \ref{tab:RMSE_FT250} (RMSE) and \ref{tab:MAE_FT250} (MAE), is consistent with the high explanatory power of the ZML just presented. This function records the lowest mean errors (RMSE $\mu \approx 0.0032$, MAE $\mu \approx 0.0013$) accompanied by small dispersions (RMSE $\sigma \approx 0.0023$, MAE $\sigma \approx 0.0008$). Moreover, the ZML attains the lowest maximum MAE and the second-lowest
maximum RMSE, after the Exp, indicating good control over
worst-case fitting errors.
Interestingly, the Exp achieves lower error metrics than the Power and ShPower, as shown in Tables~\ref{tab:RMSE_FT250} and~\ref{tab:MAE_FT250}. However, its lower $R^2$ suggests that purely exponential decay explains less of the data's total variance than the heavy-tailed laws.

\noindent
Table~\ref{tab:AIC_FT250} reports the AIC results, which account for the
number of estimated parameters. The ZML has the lowest mean
AIC ($\mu\approx-2003$) and median AIC ($Me\approx-1953$) among the candidate
functions and therefore ranks first according to this criterion in the FTSE
250 sample.

\begin{table}[!htbp]
  \centering
  \begin{minipage}{0.45\textwidth}
    \caption{Summary statistics of the pooled $R^2$ values for the FTSE 250 dataset (higher values indicate better fit).}
    \resizebox{\textwidth}{!}{ 
    \begin{tabular}{|c|c|c|c|c|c|}
    \toprule
    $R^2$    & $\mu$  & $\sigma$   & $Me$ & $\min$   & $\max$ \\
    \midrule
    Exp & \cellcolor[rgb]{ .992,  .831,  .498}0.8865 & \cellcolor[rgb]{ .988,  .663,  .471}0.1371 & \cellcolor[rgb]{ .996,  .906,  .514}0.9560 & \cellcolor[rgb]{ .984,  .667,  .467}0.4341 & \cellcolor[rgb]{ .996,  .914,  .514}0.9986 \\
    \midrule
    Power & \cellcolor[rgb]{ .996,  .878,  .506}0.9201 & \cellcolor[rgb]{ 1,  .89,  .514}0.0634 & \cellcolor[rgb]{ .996,  .875,  .506}0.9339 & \cellcolor[rgb]{ 1,  .922,  .518}0.7088 & \cellcolor[rgb]{ .996,  .918,  .514}0.9991 \\
    \midrule
    ShPower & \cellcolor[rgb]{ 1,  .922,  .518}0.9501 & \cellcolor[rgb]{ .78,  .855,  .502}0.0431 & \cellcolor[rgb]{ 1,  .922,  .518}0.9646 & \cellcolor[rgb]{ .475,  .773,  .49}0.7735 & \cellcolor[rgb]{ .937,  .906,  .518}0.9992 \\
    \midrule
    DGBD  & \cellcolor[rgb]{ .98,  .918,  .518}0.9509 & \cellcolor[rgb]{ .788,  .859,  .502}0.0435 & \cellcolor[rgb]{ .965,  .914,  .518}0.9655 & \cellcolor[rgb]{ .412,  .753,  .486}0.7810 & \cellcolor[rgb]{ 1,  .922,  .518}0.9991 \\
    \midrule
    ZML & \cellcolor[rgb]{ .388,  .745,  .482}0.9724 & \cellcolor[rgb]{ .388,  .745,  .482}0.0273 & \cellcolor[rgb]{ .388,  .745,  .482}0.9803 & \cellcolor[rgb]{ .388,  .745,  .482}0.7839 & \cellcolor[rgb]{ .388,  .745,  .482}0.9996 \\
    \midrule
    Poly & \cellcolor[rgb]{ .973,  .412,  .42}0.5836 & \cellcolor[rgb]{ .973,  .412,  .42}0.2190 & \cellcolor[rgb]{ .973,  .412,  .42}0.6192 & \cellcolor[rgb]{ .973,  .412,  .42}0.1554 & \cellcolor[rgb]{ .973,  .412,  .42}0.9616 \\
    \midrule
    Uni & \cellcolor[rgb]{ .835,  .875,  .51}0.9562 & \cellcolor[rgb]{ 1,  .922,  .518}0.0519 & \cellcolor[rgb]{ .761,  .855,  .506}0.9708 & \cellcolor[rgb]{ .984,  .678,  .471}0.4465 & \cellcolor[rgb]{ .635,  .816,  .498}0.9994 \\
    \bottomrule
    \end{tabular}%
    }
    \label{tab:R2_FT250}%
  \end{minipage}
  \hfill
  \begin{minipage}{0.45\textwidth}
    \caption{Summary statistics of the pooled RMSE values for the FTSE 250 dataset (lower values indicate better fit).}
  \resizebox{\textwidth}{!}{ 
   \begin{tabular}{|c|c|c|c|c|c|}
   \toprule
    RMSE   & $\mu$  & $\sigma$   & $Me$ & $\min$   & $\max$ \\
    \midrule
    Exp & \cellcolor[rgb]{ .953,  .906,  .514}0.0044 & \cellcolor[rgb]{ .388,  .745,  .482}0.0020 & \cellcolor[rgb]{ 1,  .914,  .518}0.0040 & \cellcolor[rgb]{ .973,  .412,  .42}0.0012 & \cellcolor[rgb]{ .388,  .745,  .482}0.0152 \\
    \midrule
    Power & \cellcolor[rgb]{ 1,  .875,  .51}0.0056 & \cellcolor[rgb]{ 1,  .906,  .518}0.0038 & \cellcolor[rgb]{ .996,  .82,  .498}0.0054 & \cellcolor[rgb]{ 1,  .902,  .518}0.0004 & \cellcolor[rgb]{ 1,  .922,  .518}0.0194 \\
    \midrule
    ShPower & \cellcolor[rgb]{ 1,  .918,  .518}0.0046 & \cellcolor[rgb]{ 1,  .918,  .518}0.0036 & \cellcolor[rgb]{ 1,  .922,  .518}0.0039 & \cellcolor[rgb]{ .733,  .843,  .502}0.0004 & \cellcolor[rgb]{ 1,  .922,  .518}0.0193 \\
    \midrule
    DGBD  & \cellcolor[rgb]{ 1,  .922,  .518}0.0045 & \cellcolor[rgb]{ 1,  .922,  .518}0.0035 & \cellcolor[rgb]{ .929,  .898,  .51}0.0037 & \cellcolor[rgb]{ .608,  .808,  .494}0.0004 & \cellcolor[rgb]{ .992,  .918,  .514}0.0193 \\
    \midrule
    ZML & \cellcolor[rgb]{ .388,  .745,  .482}0.0032 & \cellcolor[rgb]{ .506,  .776,  .486}0.0023 & \cellcolor[rgb]{ .388,  .745,  .482}0.0028 & \cellcolor[rgb]{ 1,  .922,  .518}0.0004 & \cellcolor[rgb]{ .498,  .776,  .486}0.0159 \\
    \midrule
    Poly & \cellcolor[rgb]{ .973,  .412,  .42}0.0156 & \cellcolor[rgb]{ .973,  .412,  .42}0.0120 & \cellcolor[rgb]{ .973,  .412,  .42}0.0116 & \cellcolor[rgb]{ .984,  .576,  .451}0.0009 & \cellcolor[rgb]{ .973,  .412,  .42}0.0516 \\
    \midrule
    Uni & \cellcolor[rgb]{ .733,  .843,  .502}0.0039 & \cellcolor[rgb]{ .816,  .867,  .506}0.0031 & \cellcolor[rgb]{ .675,  .827,  .498}0.0033 & \cellcolor[rgb]{ .388,  .745,  .482}0.0003 & \cellcolor[rgb]{ 1,  .902,  .514}0.0206 \\
    \bottomrule
    \end{tabular}%
    }
    \label{tab:RMSE_FT250}%
  \end{minipage}
\end{table}

\begin{table}[!htbp]
  \centering
  \begin{minipage}{0.45\textwidth}
   \caption{Summary statistics of the pooled MAE values for the FTSE 250
dataset (lower values indicate better fit).}
    \resizebox{\textwidth}{!}{ 
    \begin{tabular}{|c|c|c|c|c|c|}
    \toprule
    MAE   & $\mu$  & $\sigma$   & $Me$ & $\min$   & $\max$ \\
    \midrule
    Exp & \cellcolor[rgb]{ .714,  .839,  .498}0.0016 & \cellcolor[rgb]{ .447,  .761,  .482}0.0008 & \cellcolor[rgb]{ .624,  .812,  .494}0.0015 & \cellcolor[rgb]{ .992,  .757,  .486}0.0004 & \cellcolor[rgb]{ .608,  .808,  .494}0.0052 \\
    \midrule
    Power & \cellcolor[rgb]{ .996,  .8,  .494}0.0033 & \cellcolor[rgb]{ .996,  .808,  .498}0.0020 & \cellcolor[rgb]{ .992,  .725,  .482}0.0037 & \cellcolor[rgb]{ 1,  .91,  .518}0.0002 & \cellcolor[rgb]{ 1,  .906,  .518}0.0075 \\
    \midrule
    ShPower & \cellcolor[rgb]{ 1,  .871,  .51}0.0025 & \cellcolor[rgb]{ 1,  .863,  .506}0.0017 & \cellcolor[rgb]{ 1,  .851,  .506}0.0025 & \cellcolor[rgb]{ .388,  .745,  .482}0.0002 & \cellcolor[rgb]{ 1,  .922,  .518}0.0071 \\
    \midrule
    DGBD  & \cellcolor[rgb]{ 1,  .922,  .518}0.0019 & \cellcolor[rgb]{ 1,  .922,  .518}0.0012 & \cellcolor[rgb]{ 1,  .922,  .518}0.0018 & \cellcolor[rgb]{ .612,  .808,  .494}0.0002 & \cellcolor[rgb]{ .706,  .835,  .498}0.0057 \\
    \midrule
    ZML & \cellcolor[rgb]{ .388,  .745,  .482}0.0013 & \cellcolor[rgb]{ .388,  .745,  .482}0.0008 & \cellcolor[rgb]{ .388,  .745,  .482}0.0012 & \cellcolor[rgb]{ 1,  .922,  .518}0.0002 & \cellcolor[rgb]{ .388,  .745,  .482}0.0041 \\
    \midrule
    Poly & \cellcolor[rgb]{ .973,  .412,  .42}0.0074 & \cellcolor[rgb]{ .973,  .412,  .42}0.0049 & \cellcolor[rgb]{ .973,  .412,  .42}0.0068 & \cellcolor[rgb]{ .973,  .412,  .42}0.0006 & \cellcolor[rgb]{ .973,  .412,  .42}0.0166 \\
    \midrule
    Uni & \cellcolor[rgb]{ .82,  .867,  .506}0.0017 & \cellcolor[rgb]{ .827,  .871,  .506}0.0011 & \cellcolor[rgb]{ .839,  .875,  .506}0.0017 & \cellcolor[rgb]{ .553,  .792,  .49}0.0002 & \cellcolor[rgb]{ .996,  .784,  .494}0.0097 \\
    \bottomrule
    \end{tabular}%
    }
    \label{tab:MAE_FT250}%
  \end{minipage}
  \hfill
  \begin{minipage}{0.45\textwidth}
    \caption{Summary statistics of the pooled AIC values for the FTSE 250 dataset (lower values indicate better fit).}
  \resizebox{\textwidth}{!}{ 
   \begin{tabular}{|c|c|c|c|c|c|}
    \toprule
    AIC   & $\mu$  & $\sigma$   & $Me$ & $\min$   & $\max$ \\
    \midrule
    Exp & \cellcolor[rgb]{ .996,  .827,  .502}-1832.96 & \cellcolor[rgb]{ .388,  .745,  .482}152.83 & \cellcolor[rgb]{ 1,  .906,  .518}-1827.91 & \cellcolor[rgb]{ .973,  .412,  .42}-2241.66 & \cellcolor[rgb]{ .388,  .745,  .482}-1388.80 \\
    \midrule
    Power & \cellcolor[rgb]{ .996,  .835,  .502}-1838.76 & \cellcolor[rgb]{ .988,  .694,  .475}322.81 & \cellcolor[rgb]{ .992,  .769,  .49}-1729.04 & \cellcolor[rgb]{ 1,  .89,  .514}-2586.89 & \cellcolor[rgb]{ 1,  .922,  .518}-1306.23 \\
    \midrule
    ShPower & \cellcolor[rgb]{ 1,  .922,  .518}-1908.05 & \cellcolor[rgb]{ .988,  .671,  .471}324.07 & \cellcolor[rgb]{ 1,  .922,  .518}-1840.77 & \cellcolor[rgb]{ .729,  .843,  .502}-2627.14 & \cellcolor[rgb]{ 1,  .922,  .518}-1307.46 \\
    \midrule
    DGBD  & \cellcolor[rgb]{ .988,  .918,  .514}-1909.81 & \cellcolor[rgb]{ 1,  .922,  .518}311.50 & \cellcolor[rgb]{ .945,  .906,  .514}-1850.75 & \cellcolor[rgb]{ .596,  .804,  .494}-2634.72 & \cellcolor[rgb]{ .992,  .918,  .514}-1308.20 \\
    \midrule
    ZML & \cellcolor[rgb]{ .388,  .745,  .482}-2003.18 & \cellcolor[rgb]{ .867,  .882,  .51}277.15 & \cellcolor[rgb]{ .388,  .745,  .482}-1953.21 & \cellcolor[rgb]{ 1,  .922,  .518}-2611.81 & \cellcolor[rgb]{ .514,  .78,  .486}-1371.62 \\
    \midrule
    Poly & \cellcolor[rgb]{ .973,  .412,  .42}-1514.67 & \cellcolor[rgb]{ .973,  .412,  .42}336.78 & \cellcolor[rgb]{ .973,  .412,  .42}-1476.08 & \cellcolor[rgb]{ .98,  .522,  .443}-2319.36 & \cellcolor[rgb]{ .973,  .412,  .42}-979.87 \\
    \midrule
    Uni & \cellcolor[rgb]{ .843,  .875,  .506}-1932.03 & \cellcolor[rgb]{ .847,  .875,  .506}272.03 & \cellcolor[rgb]{ .714,  .835,  .498}-1893.31 & \cellcolor[rgb]{ .388,  .745,  .482}-2646.59 & \cellcolor[rgb]{ 1,  .886,  .514}-1283.28 \\
    \bottomrule
    \end{tabular}%
    }
    \label{tab:AIC_FT250}%
  \end{minipage}\end{table}

The summary measures do not rank the candidate functions uniformly across all
distributional features. We therefore apply the stochastic-dominance
procedure described in Section~\ref{Sec:evaluation} to the empirical
distribution of RMSE. We use RMSE because the rank-size parameters are
estimated by least squares, so this measure is directly linked to the fitting
criterion and penalizes large errors.

\begin{figure}[!htbp]
    \centering
        \includegraphics[width=0.61\linewidth]{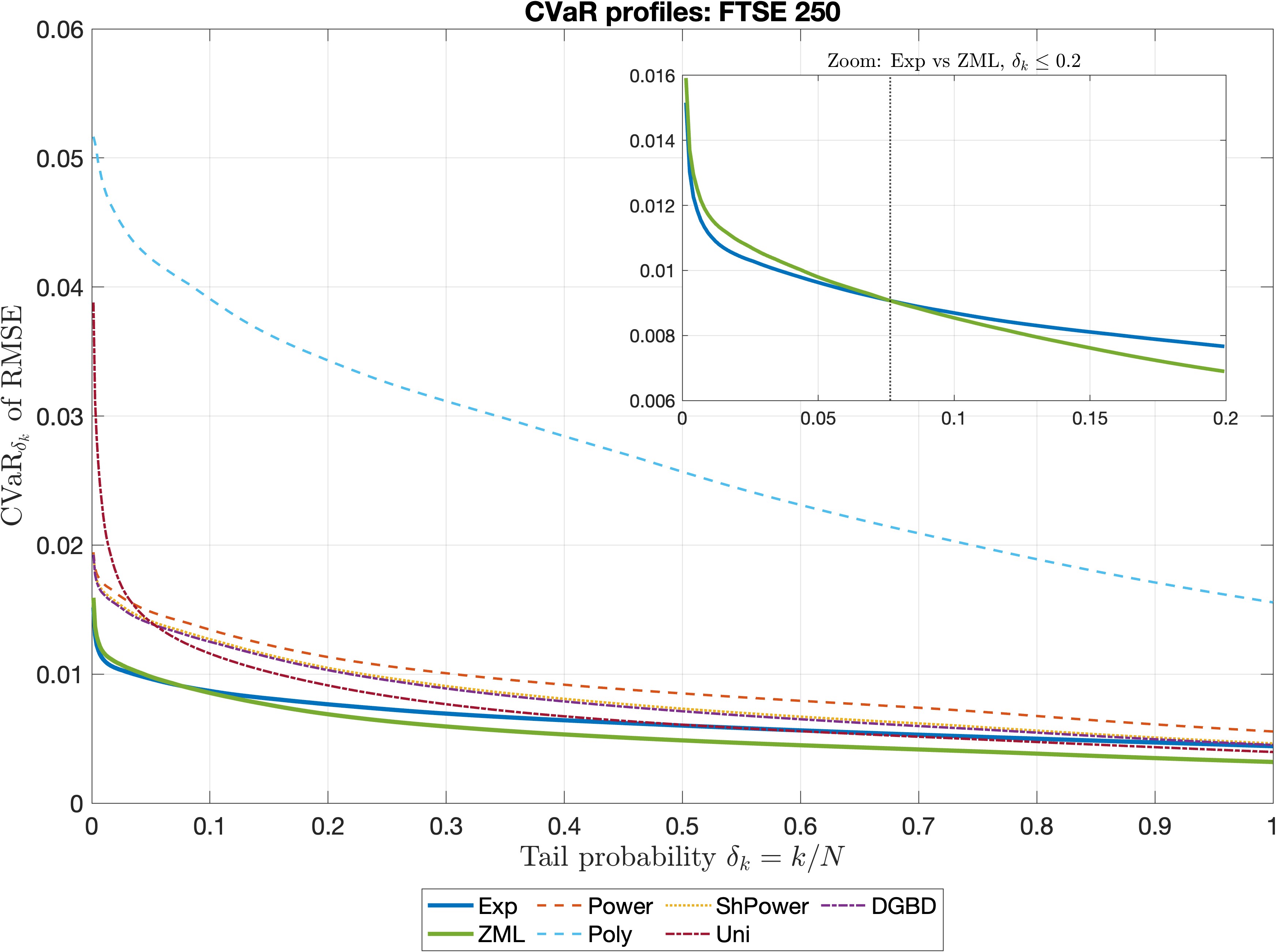}
        \caption{Empirical CVaR profiles computed from the pooled RMSE distribution
of each of the seven candidate rank-size functions for the FTSE 250 dataset.
For each function, the empirical distribution contains
$N=MQ=4\times183=732$ RMSE observations pooled across $M=4$
portfolio-selection models and $Q=183$ rolling windows. At each tail
probability $\delta_k=k/N$, $k=1,\ldots,N$, lower CVaR values indicate better tail-fitting
performance because they correspond to a smaller average RMSE among the $k$ worst realizations.}
        \label{fig:CVaR_stacked_FTSE}
\end{figure}

\noindent
Figure~\ref{fig:CVaR_stacked_FTSE} reports the empirical
$\operatorname{CVaR}_{\delta_k}$ profiles of the pooled RMSE values, where $\delta_k=k/N$, $k=1,\ldots, N$, is the fraction of the worst observations included in the tail average. Lower values indicate better fitting performance.

\noindent
For tail probabilities up to $\delta_k\approx0.075$, the Exp has the lowest empirical CVaR.
For larger tail probabilities, the ZML has the lowest values.
The crossing of the two profiles implies that neither function SSD-dominates
the other. We therefore use CS$\varepsilon$SD to measure the relaxation
required in each direction.

\noindent
To determine the robust winner between the two leading candidates, namely, the Exp and the ZML, we apply the procedure described in Section \ref{Sec:evaluation}.
Specifically, for each directional comparison, \(X\) against \(Y\)
and \(Y\) against \(X\), we compute the CVaR differences at each tail
probability level $\delta_k=k/N$, with $N=732$. For the comparison of \(X\) with \(Y\), these
differences are given by
\[
d_k(X,Y)
=
\operatorname{CVaR}_{\delta_k}(X)
-
\operatorname{CVaR}_{\delta_k}(Y),
\qquad k=1,\ldots, N.
\]
We arrange them in non-increasing order and compute the corresponding
cumulative sums
\[
C_t(X,Y)
=
\sum_{j=1}^{t}d^{(j)}(X,Y),
\qquad t=1,\ldots, N.
\]
The maximum of these cumulative sums is the signed dominance index
\[
\mathcal{D}(X,Y)
=
\max_{t=1,\ldots, N}C_t(X,Y),
\]
as defined in Eq.~\eqref{eq:signed_dominance_index}. We repeat the
same procedure in the opposite direction to obtain
\(\mathcal{D}(Y,X)\). Comparing the two indices allows us to determine
which rank-size function exhibits the stronger dominance relation, or
requires the smaller relaxation when exact SSD does not hold.

\begin{figure}[!htbp]
    \centering
    \begin{minipage}[t]{0.49\linewidth}
        \centering
         \includegraphics[width=\linewidth]{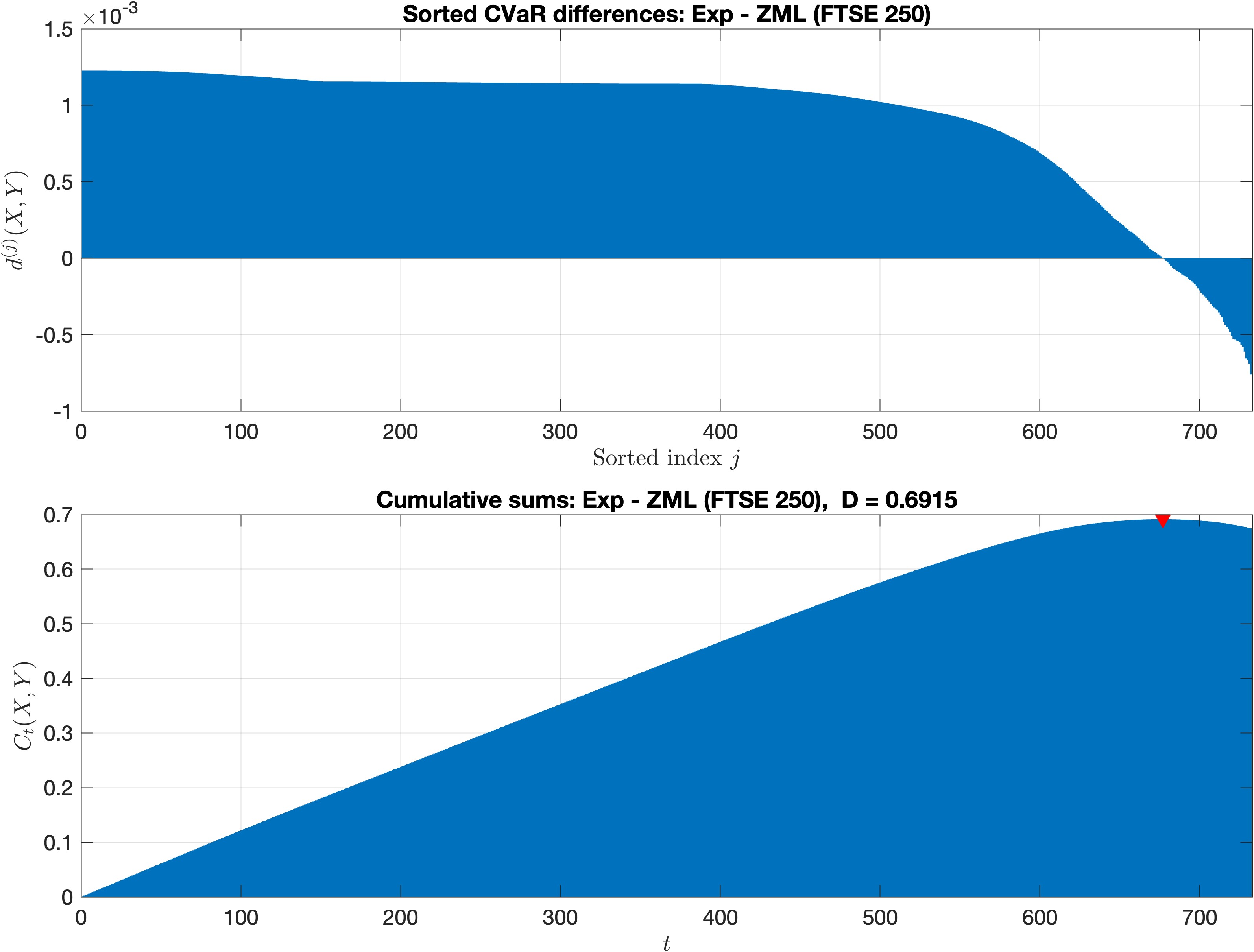}
    \caption{CVaR differences Exp-ZML (all $N=732$ tail levels)}
    \label{fig:CVaR_Diff_exp_Zipf_FT250}
    \end{minipage}
    \hfill
    \begin{minipage}[t]{0.49\linewidth}
        \centering
         \includegraphics[width=\linewidth]{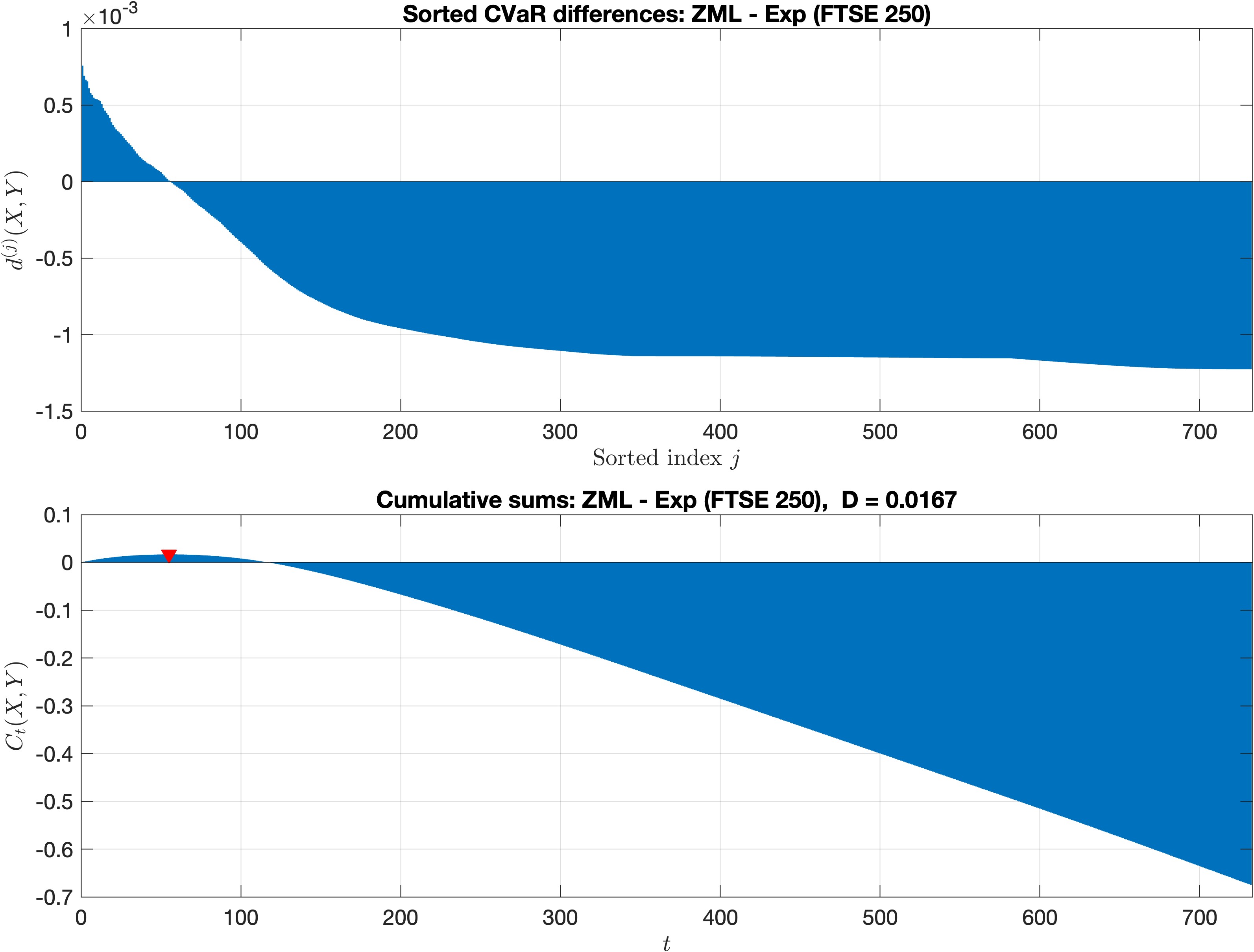}
    \caption{CVaR differences ZML-Exp (all $N=732$ tail levels)}
    \label{fig:CVaR_Diff_Zipf_exp_FT250}
    \end{minipage}
\end{figure}

Figures \ref{fig:CVaR_Diff_exp_Zipf_FT250} and \ref{fig:CVaR_Diff_Zipf_exp_FT250} illustrate this process for the two opposing directions.
\begin{itemize}
    \item When testing whether the Exp dominates the ZML with respect to CS\(\varepsilon\)SD (Figure \ref{fig:CVaR_Diff_exp_Zipf_FT250}), the cumulative sums of the ordered differences reach a maximum signed dominance index of
\[
\mathcal{D}(X^{(\mathrm{Exp})},X^{(\mathrm{ZML})})
= 0.6915.
\]
Since this value is positive, exact SSD does not hold and the minimum
required tolerance is
\[
\varepsilon_{\min}
\bigl(X^{(\mathrm{Exp})},X^{(\mathrm{ZML})}\bigr)
= 0.6915.
\]
    
    \item Conversely, when testing whether the ZML dominates the
Exp (Figure \ref{fig:CVaR_Diff_Zipf_exp_FT250}), the maximum cumulative
sum is
\[
\mathcal{D}(X^{(\mathrm{ZML})},X^{(\mathrm{Exp})})
= 0.0167.
\]
Thus, exact SSD does not hold in this direction either, but only a
small relaxation is required:
\[
\varepsilon_{\min}
\bigl(X^{(\mathrm{ZML})},X^{(\mathrm{Exp})}\bigr)
= 0.0167.
\]
\end{itemize}

\noindent
Since
\[
\mathcal{D}
\bigl(X^{(\mathrm{ZML})},X^{(\mathrm{Exp})}\bigr)
<
\mathcal{D}
\bigl(X^{(\mathrm{Exp})},X^{(\mathrm{ZML})}\bigr),
\]
and the corresponding minimum tolerances are \(0.0167\) and \(0.6915\), respectively; the pairwise comparison clearly favors the ZML.

\noindent
To extend the analysis to all candidate functions, Table \ref{tab:pairwise_dominance_FT} presents the complete pairwise comparison matrix. 
Each cell \((i,j)\) reports the signed dominance index
\[
\mathcal{D}\bigl(X^{(i)},X^{(j)}\bigr),
\]
where \(X^{(i)}\) and \(X^{(j)}\) are the fitting-performance
variables associated with the rank-size functions displayed in row
\(i\) and column \(j\), respectively. A non-positive value indicates
that the row function weakly SSD-dominates the column function and
therefore requires no positive relaxation. In particular, a negative
value indicates that all the corresponding CVaR inequalities are
strict. A positive value indicates that exact SSD does not hold; in
this case, the reported value coincides with the minimum tolerance
required for the row function to CS\(\varepsilon\)SD-dominate the
column function.
The robustness of a rank-size function is therefore identified by looking for a row with consistently low values (indicating it easily dominates others) and a column with consistently high values (indicating it is difficult to be dominated by others).

\begin{table}[htbp]
  \centering
   \resizebox{0.8\textwidth}{!}{ 
    \begin{tabular}{|c|c|c|c|c|c|c|c|}
\cmidrule{2-8}    \multicolumn{1}{r|}{} & Exp & Power & ShPower & DGBD  & ZML & Poly & Uni \\
    \midrule
    Exp & -      & \cellcolor[rgb]{ .726,  .843,  .502}-0.00114 & \cellcolor[rgb]{ .4,  .749,  .483}-0.00021 & \cellcolor[rgb]{ .394,  .747,  .482}-0.00010 & \cellcolor[rgb]{ .388,  .745,  .482}0.69146 & \cellcolor[rgb]{ .977,  .915,  .517}-0.01114 & \cellcolor[rgb]{ .456,  .765,  .486}0.07730 \\
    \midrule
    Power & \cellcolor[rgb]{ .998,  .892,  .512}2.02586 & -      & \cellcolor[rgb]{ .999,  .896,  .513}0.74705 & \cellcolor[rgb]{ .999,  .894,  .513}0.88146 & \cellcolor[rgb]{ .998,  .892,  .512}2.70064 & \cellcolor[rgb]{ .973,  .412,  .42}-0.01001 & \cellcolor[rgb]{ .998,  .893,  .512}1.49715 \\
    \midrule
    ShPower & \cellcolor[rgb]{ 1,  .92,  .518}1.27881 & \cellcolor[rgb]{ 1,  .922,  .518}-0.00013 & -      & \cellcolor[rgb]{ 1,  .922,  .518}0.13441 & \cellcolor[rgb]{ 1,  .92,  .518}1.95359 & \cellcolor[rgb]{ .996,  .844,  .503}-0.01093 & \cellcolor[rgb]{ 1,  .92,  .518}0.76843 \\
    \midrule
    DGBD & \cellcolor[rgb]{ .966,  .912,  .516}1.14440 & \cellcolor[rgb]{ .994,  .92,  .518}-0.00017 & \cellcolor[rgb]{ .402,  .749,  .483}-0.00004 & -      & \cellcolor[rgb]{ .966,  .912,  .516}1.81918 & \cellcolor[rgb]{ .999,  .899,  .514}-0.01105 & \cellcolor[rgb]{ .945,  .906,  .515}0.64128 \\
    \midrule
    ZML & \cellcolor[rgb]{ .388,  .745,  .482}0.01667 & \cellcolor[rgb]{ .388,  .745,  .482}-0.00236 & \cellcolor[rgb]{ .388,  .745,  .482}-0.00144 & \cellcolor[rgb]{ .388,  .745,  .482}-0.00132 & -      & \cellcolor[rgb]{ .388,  .745,  .482}-0.01237 & \cellcolor[rgb]{ .388,  .745,  .482}-0.00078 \\
    \midrule
    Poly & \cellcolor[rgb]{ .973,  .412,  .42}15.03198 & \cellcolor[rgb]{ .973,  .412,  .42}13.00612 & \cellcolor[rgb]{ .973,  .412,  .42}13.75317 & \cellcolor[rgb]{ .973,  .412,  .42}13.88758 & \cellcolor[rgb]{ .973,  .412,  .42}15.70676 & -      & \cellcolor[rgb]{ .973,  .412,  .42}14.40391 \\
    \midrule
    Uni & \cellcolor[rgb]{ .741,  .847,  .503}0.70536 & \cellcolor[rgb]{ 1,  .918,  .517}0.09936 & \cellcolor[rgb]{ 1,  .92,  .518}0.11768 & \cellcolor[rgb]{ .978,  .916,  .517}0.12495 & \cellcolor[rgb]{ .701,  .836,  .5}1.30285 & \cellcolor[rgb]{ .761,  .853,  .504}-0.01159 & -      \\
    \bottomrule
    \end{tabular}%
    }
\caption{Pairwise signed dominance indices for the FTSE dataset.
Each entry \((i,j)\) reports
\(\mathcal{D}(X^{(i)},X^{(j)})\), where the row function is compared
with the column function. Row and column labels denote the rank-size
functions, abbreviated as in Table~\ref{tab:rank-size_fun}. The indices are computed on all the $N=732$ tail levels $\delta_k=k/N$.}
  \label{tab:pairwise_dominance_FT}%
\end{table}%

Focusing on the ZML row, we observe that the signed dominance indices are non-positive against most competing functions. For example,
$
\mathcal{D}
\bigl(X^{(\mathrm{ZML})},X^{(\mathrm{ShPower})}\bigr)
=
-0.0014,
$
which indicates that the ZML satisfies all the corresponding SSD
inequalities without requiring any positive relaxation. In the
comparison with the Exp, the index is slightly
positive:
$
\mathcal{D}
\bigl(X^{(\mathrm{ZML})},X^{(\mathrm{Exp})}\bigr)
=
0.0167.
$
Thus, the ZML requires only a small tolerance to dominate the Exp.
In the opposite direction, however, the Exp
requires a substantially larger tolerance:
$
\mathcal{D}
\bigl(X^{(\mathrm{Exp})},X^{(\mathrm{ZML})}\bigr)
=
0.6915.
$
The strong asymmetry between the two directional comparisons favors
the ZML as the most robust rank-size specification for the FTSE
dataset.
Accordingly, we adopt its calibrated parameters as the input features for the reverse engineering framework presented in the next section.

\subsection{Reverse-engineering application}\label{Sec:reverse_results}

We apply the classification framework of Sect.~\ref{Sec:reverse} to the
FTSE 250 dataset. The feature vector of each observation contains the three
calibrated ZML parameters,
$\boldsymbol{z}_{m,q}=(\widehat{E}_{m,q},\widehat{\eta}_{m,q},\widehat{\lambda}_{m,q})$,
so that $p_{f^{\star}}=3$, and the dataset contains $N=732$ observations. The
RF consists of $N_{\mathrm{tree}}=500$ trees with a minimum leaf size of three
observations. At each split, the splitting feature is selected with the
curvature test \citep{loh1997split}. All computations use a fixed random
seed.

\noindent
With $L=250$ and $s=20$, Eq.~\eqref{eq:embargo} gives an embargo of $h=12$
windows. We partition the $Q=183$ windows into $V=5$ blocks of consecutive
windows (windows 1--37, 38--73, 74--110, 111--146, and 147--183). After the
removal of the windows within the embargo, each training set contains between
122 and 134 windows, that is, between 488 and 536 observations. When the
selection procedure of Sect.~\ref{Sec:evaluation} is repeated within each
training set, the ZML is selected in all five blocks. The feature vector is
therefore the same in all blocks and does not depend on the test
observations.

\noindent
Table~\ref{tab:rf_blocked} reports the accuracy in each test block, together
with its mean, standard deviation, and $95\%$ confidence interval. Since the
classes are balanced, the chance level is $50\%$ at the strategy level and
$25\%$ at the model level.

\begin{table}[!htbp]
\centering
\caption{Accuracy (\%) of the RF classifier under blocked cross-validation
with embargo for the FTSE 250 dataset. The chance level is the accuracy of a
random assignment of labels.}
\label{tab:rf_blocked}
\resizebox{\textwidth}{!}{
\begin{tabular}{lccccccccc}
\toprule
Task & Block 1 & Block 2 & Block 3 & Block 4 & Block 5 & Mean & Std.\ dev. & $95\%$ CI & Chance level\\
\midrule
Strategy level & 96.0 & 84.0 & 87.2 & 90.3 & 87.2 & 88.9 & 4.5 & $[83.3,\ 94.5]$ & 50.0\\
Model level    & 71.6 & 63.2 & 62.8 & 61.1 & 68.2 & 65.4 & 4.4 & $[60.0,\ 70.8]$ & 25.0\\
\bottomrule
\end{tabular}
}
\end{table}

\begin{figure}[!htbp]
    \centering
        \includegraphics[width=0.6\linewidth]{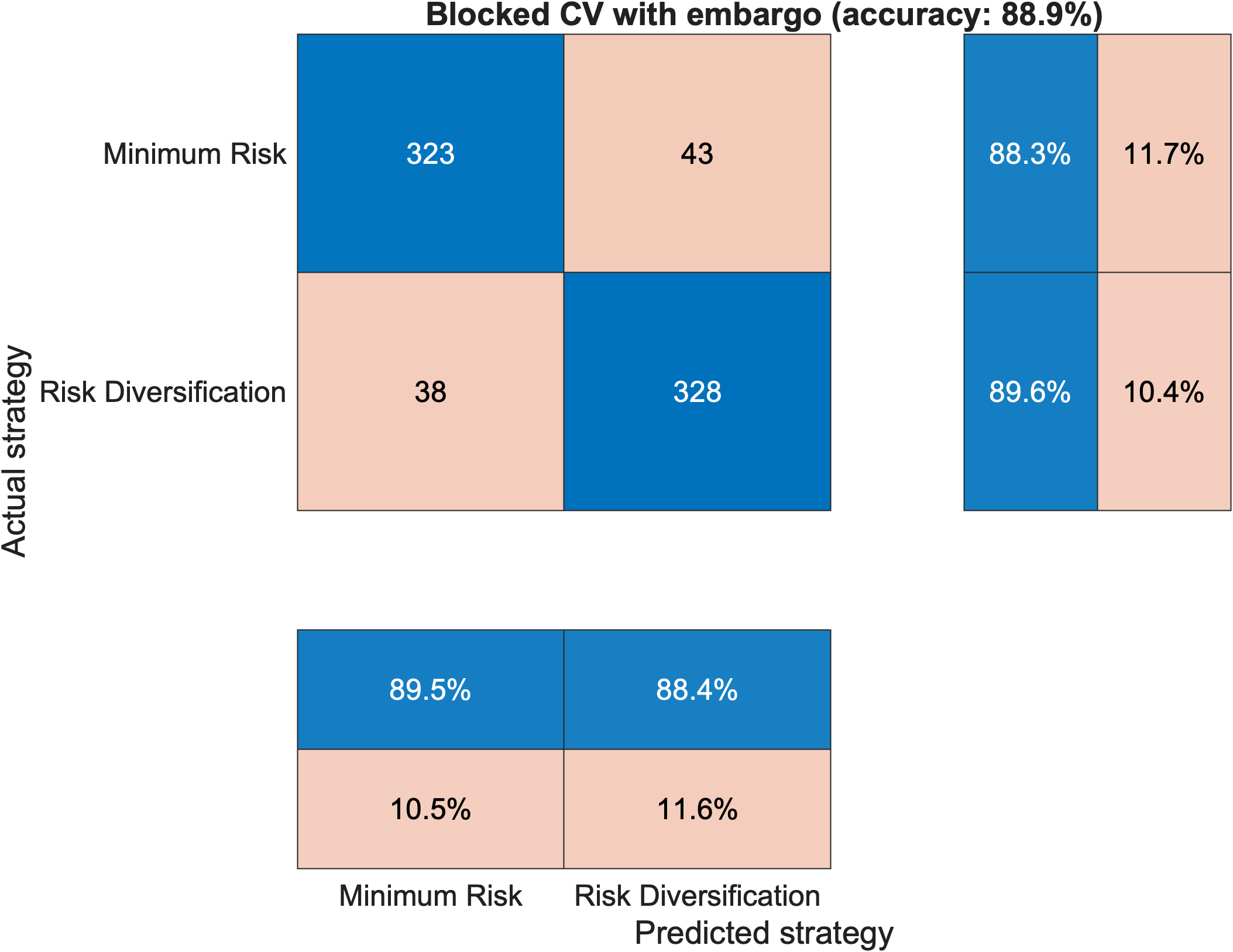}
        \caption{Pooled confusion matrix for strategy classification under
blocked cross-validation with embargo (FTSE 250). Each observation is
predicted by the classifier trained without its test block. Rows denote
actual classes and columns denote predicted classes. The row summary reports
the recall of each class and the column summary its precision.}
        \label{fig:RF_strategies}
\end{figure}

\noindent
At the strategy level, the average accuracy is $88.9\%$, with a $95\%$
confidence interval of $[83.3\%,\ 94.5\%]$. The pooled confusion matrix in
Figure~\ref{fig:RF_strategies} shows that the errors are balanced between the
two classes: the recall is $88.3\%$ for the minimum-risk class and $89.6\%$
for the risk-diversification class. Aggregating the predictions of the
model-level classifier into the two strategy classes yields a similar
accuracy ($87.8\%$).

\begin{figure}[!htbp]
    \centering
        \includegraphics[width=0.6\linewidth]{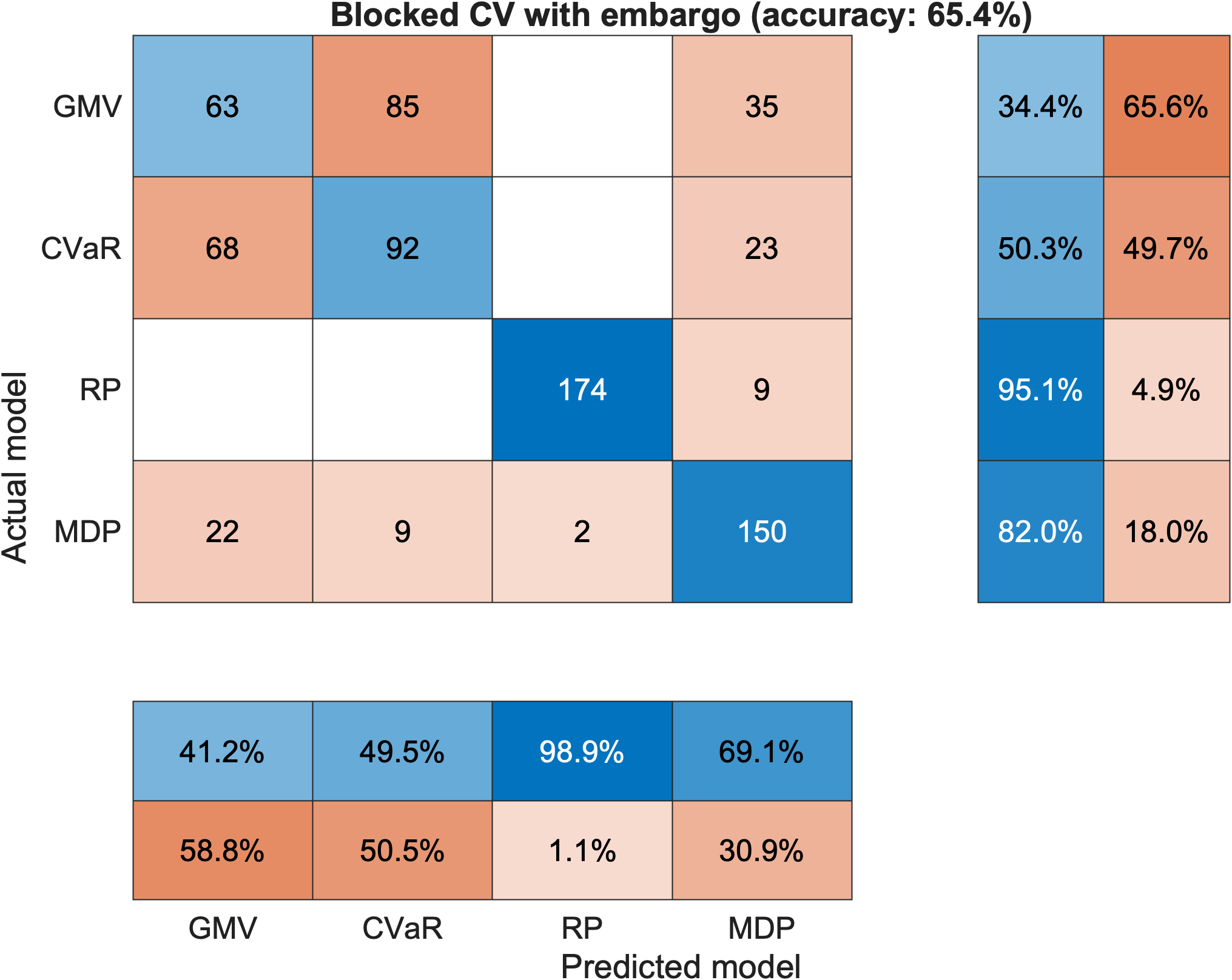}
        \caption{Pooled confusion matrix for portfolio-model classification
under blocked cross-validation with embargo (FTSE 250). Each observation is
predicted by the classifier trained without its test block. Rows denote
actual classes and columns denote predicted classes. The row summary reports
the recall of each class and the column summary its precision.}
        \label{fig:RF_models}
\end{figure}

\noindent
At the model level, the average accuracy is $65.4\%$, against a chance level
of $25\%$. Figure~\ref{fig:RF_models} shows that this result differs across
models. RP is almost always identified correctly (recall $95.1\%$), and its
few errors are classified as MDP. MDP has a recall of $82.0\%$. By contrast,
GMV and minimum-CVaR are difficult to distinguish: 85 of the 183 GMV
portfolios are classified as minimum-CVaR, and 68 of the 183 minimum-CVaR
portfolios as GMV, which gives recall rates of $34.4\%$ and $50.3\%$,
respectively. A smaller number of GMV and minimum-CVaR portfolios (35 and 23)
are classified as MDP. Most errors therefore occur within the minimum-risk
class, which explains the higher accuracy at the strategy level.

\noindent
The accuracy also varies across blocks. The recall of MDP exceeds $94\%$ in
the first three blocks and falls to $66.7\%$ and $54.1\%$ in the last two,
whereas the recall of RP falls to $75.0\%$ in block 4. The relation between
the parameters and the generating models is therefore not constant over time,
which supports a validation design that tests the classifier on periods not
used for training.

\noindent
We also apply the robustness checks described in Sect.~\ref{Sec:reverse}.
In the walk-forward validation, the test blocks consist of 18 consecutive
windows. Walk-forward validation gives accuracies similar to those of blocked
cross-validation: $63.8\%$ at the model level and $88.5\%$ at the strategy
level. A random holdout gives $72.9\%$ and $92.5\%$. 
The higher accuracy under random holdout is consistent with an optimistic bias arising from temporal dependence among overlapping estimation windows.

\noindent
Overall, the calibrated ZML parameters contain information on the generating
portfolio model that transfers to periods not used for training. They
separate minimum-risk from risk-diversification strategies and identify the
RP and MDP portfolios, whereas they do not reliably distinguish between the
two minimum-risk models. Rank-size geometry may therefore provide a useful
diagnostic input for classifying allocation rules, complementing conventional
risk--return measures.


\section{Conclusions}\label{Sec:conclusions}

This paper studies whether optimal portfolio weights can be described through
rank-size laws and whether the fitted parameters contain information about the
portfolio model that generated them. We consider four portfolio models (GMV,
minimum-CVaR, RP, and MDP) and compare the rank-size structure of their sorted
weights within a common framework.
Standard goodness-of-fit measures do not rank the candidate functions
uniformly across all distributional features. We therefore propose a
selection criterion based on Cumulative Second-order
$\varepsilon$-Stochastic Dominance (CS$\varepsilon$SD), applied to the
empirical distribution of RMSE.

\noindent
For the FTSE 250 dataset, the criterion selects the Zipf--Mandelbrot law. The
results for the other equity universes select the DGBD for the DJI, the
Exponential law for the EURO STOXX 50, and the Zipf--Mandelbrot law for the
NASDAQ 100. These results show that the selected functional form is
market-specific, while the same selection procedure can be applied across
the four datasets.
For the FTSE 250 dataset, the calibrated Zipf--Mandelbrot parameters are used
as inputs to a Random Forest classifier, evaluated by blocked
cross-validation with an embargo between training and test windows. The
average accuracy is $88.9\%$ at the strategy level and $65.4\%$ at the model
level, against chance levels of $50\%$ and $25\%$.
RP and MDP have the
highest class-specific recall rates, whereas most classification errors occur
between GMV and minimum-CVaR. 
Similar results are obtained under walk-forward
validation.

\noindent
These findings indicate that the fitted rank-size parameters contain
information associated with the generating portfolio model. The confusion
between GMV and minimum-CVaR is consistent with an overlap in their fitted
parameter patterns, but the classification results alone do not establish
the economic cause of this similarity.

Overall, the results suggest that rank-size analysis can complement
traditional risk--return measures by providing a compact description of the
geometry of portfolio allocations. The fitted parameters may support the
identification and interpretation of allocation rules, but they should be
used together with other information about the portfolio construction
process.

Future research could extend this analysis to non-equity markets to assess the generalizability of rank-size representations in modeling optimal weight distributions, investigate the temporal evolution of parameters as indicators of regime shifts, and integrate explainable AI to identify the geometric features driving strategy detection.


\appendix
\section{Additional empirical results} \label{Sec: additional_results}
To validate the robustness of the proposed framework, we extended the analysis to three major stock indices: Dow Jones (DJI), EURO STOXX 50 (STOXX50E), and NASDAQ 100 (NDX).
The empirical results confirm that the CS$\varepsilon$SD criterion effectively discriminates among competing functional forms across several market structures, identifying the best fit for each specific investment universe.

\begin{figure}[!htbp]
    \centering
        \includegraphics[width=0.6\linewidth]{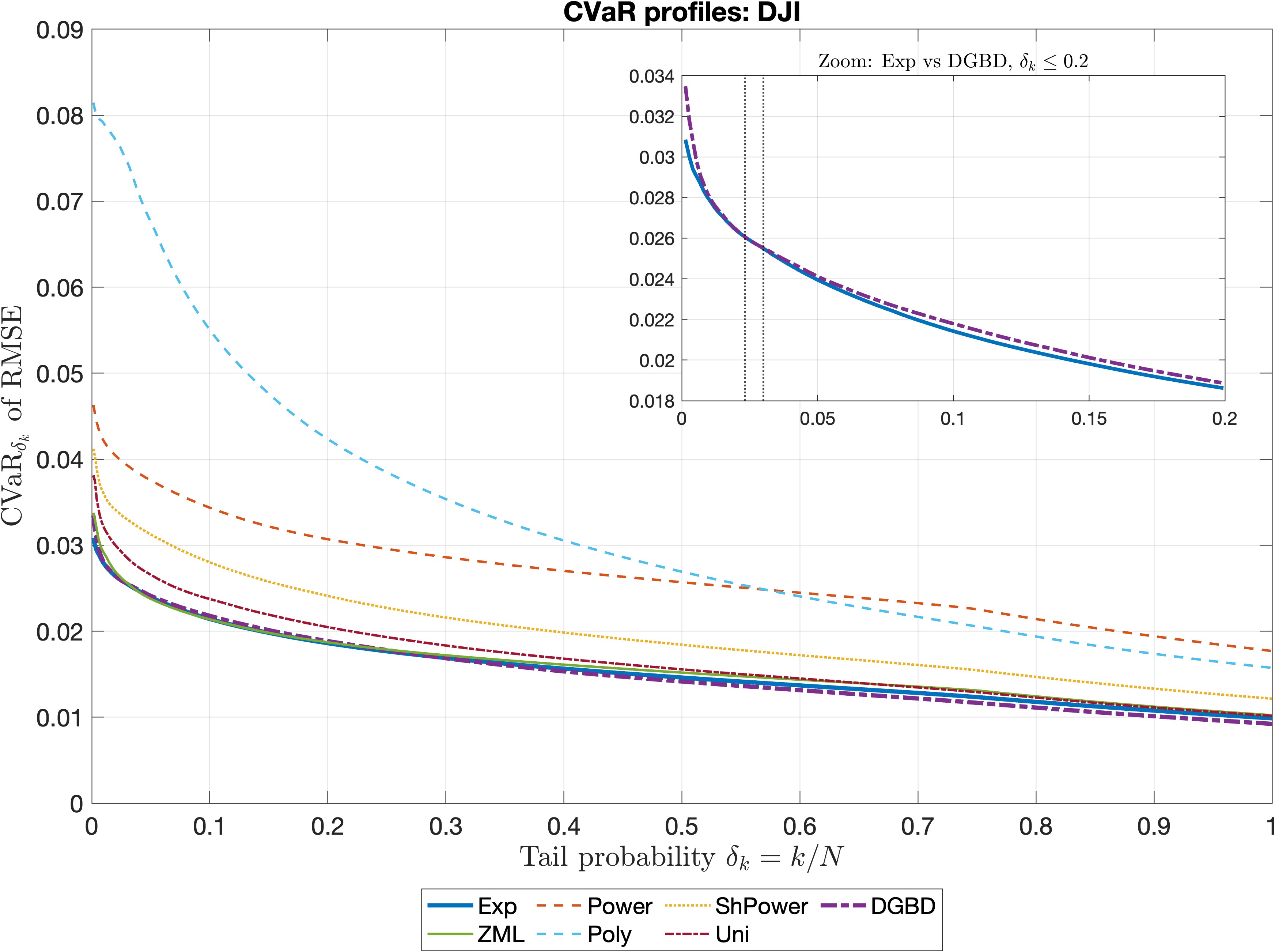}
        \caption{Empirical CVaR profiles of the pooled RMSE distributions for the
candidate rank-size functions in the Dow Jones Industrial Average dataset.
For each function, $N=MQ=4\times183=732$ RMSE observations are pooled across
$M=4$ portfolio-selection models and $Q=183$ rolling windows. At each tail
probability $\delta_k=k/N$, $k=1,\ldots,N$, lower CVaR values are preferred because they
represent smaller average RMSE among the $k$ worst realizations.}
        \label{fig:CVaR_stacked_DJI}
\end{figure}

Regarding the DJI, the DGBD emerges as the superior function, being preferred to every competing function according to Eq.~\eqref{eq:pairwise_preference}. Exact SSD does not hold against the Exp and the ZML, but the required tolerances ($0.0521$ and $0.0419$) are much smaller than those required in the opposite directions ($0.2683$ and $0.5778$) (see Figure \ref{fig:CVaR_stacked_DJI} and Table \ref{tab:pairwise_dominance_DJ}).

\begin{figure}[!htbp]
    \centering
        \includegraphics[width=0.6\linewidth]{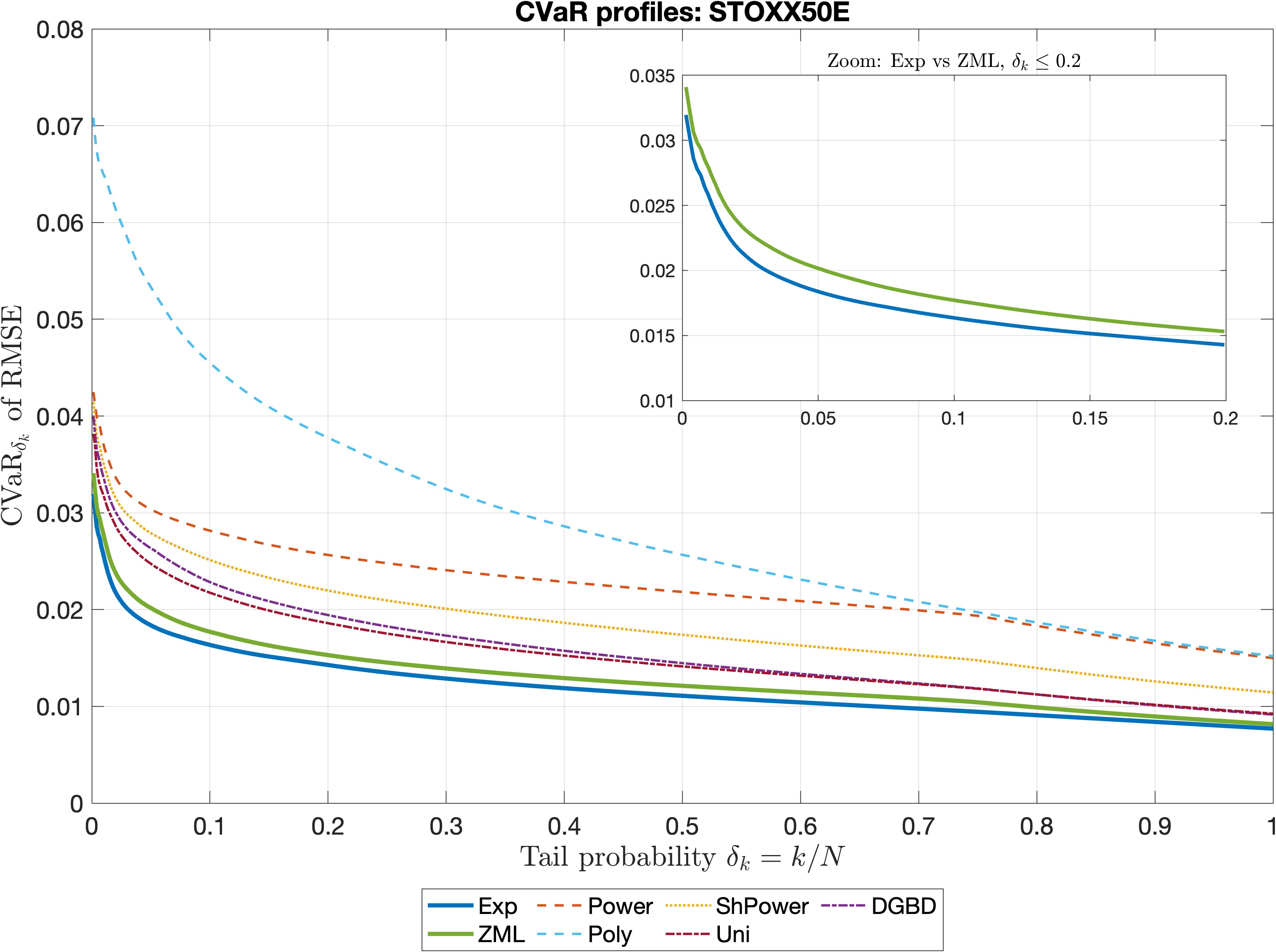}
        \caption{Empirical CVaR profiles of the pooled RMSE distributions for the
candidate rank-size functions in the EURO STOXX 50 dataset. For each
function, $N=MQ=4\times183=732$ RMSE observations are pooled across $M=4$
portfolio-selection models and $Q=183$ rolling windows. At each tail
probability $\delta_k=k/N$, $k=1,\ldots,N$, lower CVaR values are preferred because they
represent smaller average RMSE among the $k$ worst realizations.}
        \label{fig:CVaR_stacked_STOXX50E}
\end{figure}

\begin{figure}[!htbp]
    \centering
        \includegraphics[width=0.6\linewidth]{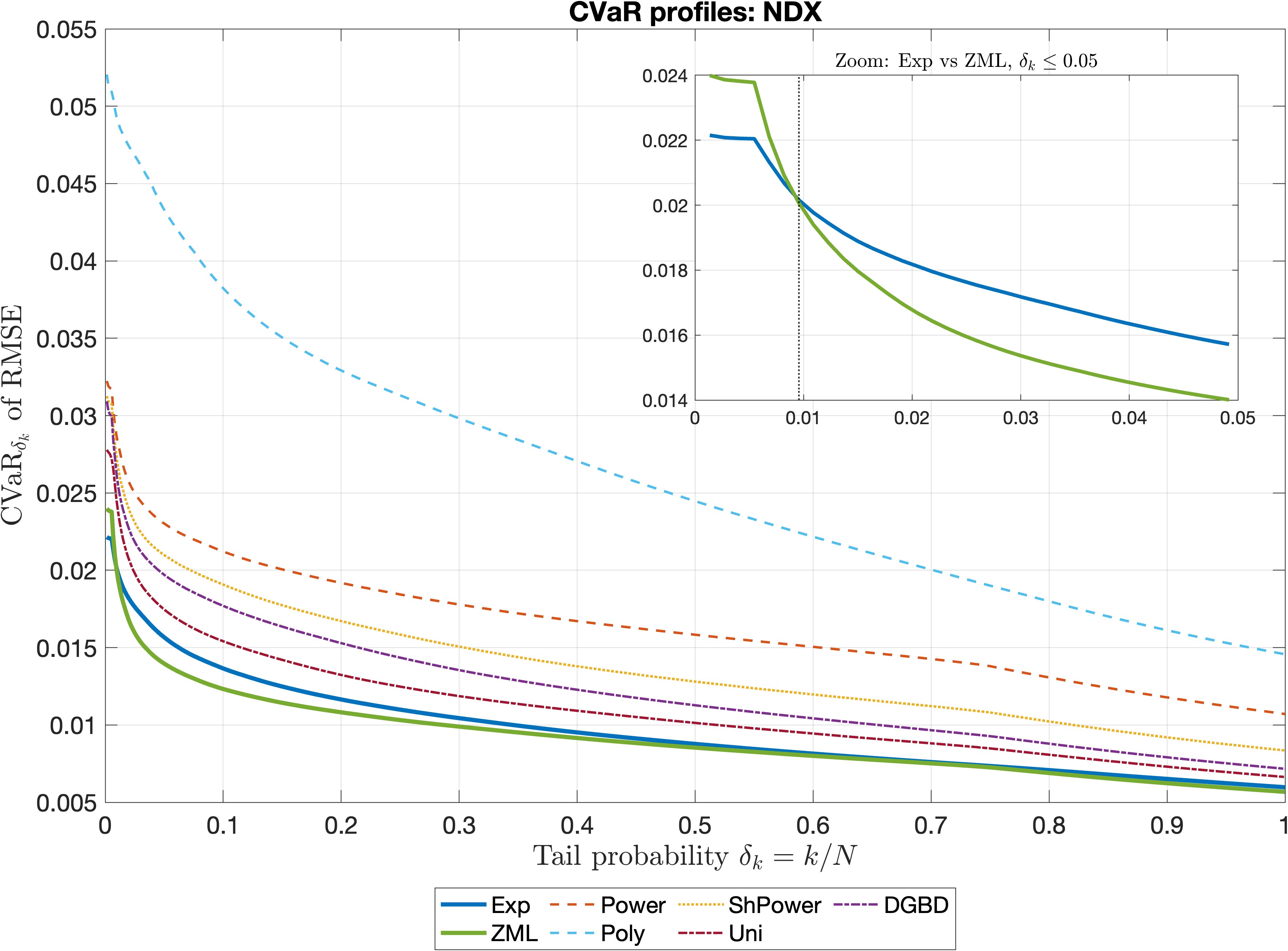}
        \caption{Empirical CVaR profiles of the pooled RMSE distributions for the
candidate rank-size functions in the NASDAQ 100 dataset. For each function,
$N=MQ=4\times183=732$ RMSE observations are pooled across $M=4$
portfolio-selection models and $Q=183$ rolling windows. At each tail
probability $\delta_k=k/N$, $k=1,\ldots,N$, lower CVaR values are preferred because they
represent smaller average RMSE among the $k$ worst realizations.}
        \label{fig:CVaR_stacked_NDX}
\end{figure}

For the STOXX50E dataset, the Exp exhibits non-positive signed dominance
indices against all competing specifications and therefore SSD-dominates
them without any positive CS\(\varepsilon\)SD relaxation (see Figure \ref{fig:CVaR_stacked_STOXX50E} and Table \ref{tab:pairwise_dominance_EX}). For the NDX dataset, the ZML SSD-dominates all competing functions except the Exp, against which only a small tolerance is required:
$\mathcal{D}\bigl(X^{(\mathrm{ZML})},X^{(\mathrm{Exp})}\bigr)=0.0081$, whereas
$\mathcal{D}\bigl(X^{(\mathrm{Exp})},X^{(\mathrm{ZML})}\bigr)=0.3494$. 
The SSD violations occur only at the six smallest tail probability levels (\(k=1,\ldots,6\)), where the ZML exhibits higher CVaR values than the Exp (see Figure \ref{fig:CVaR_stacked_NDX} and Table \ref{tab:pairwise_dominance_NDX}).
These findings indicate that, while the selected rank-size law varies across equity universes, the CS$\varepsilon$SD framework provides a common procedure for selecting among competing functional forms.

\begin{table}[!htbp]
  \centering
  \begin{minipage}{0.45\textwidth}
    \caption{Summary statistics of the pooled $R^2$ values for the DJI dataset (higher values indicate better fit).}
    \resizebox{\textwidth}{!}{ 
    \begin{tabular}{|c|c|c|c|c|c|}
    \toprule
    $R^2$    & $\mu$  & $\sigma$   & $Me$ & $\min$   & $\max$ \\
     \midrule
    Exp & \cellcolor[rgb]{ 1,  .922,  .518}0.9563 & \cellcolor[rgb]{ 1,  .922,  .518}0.0282 & \cellcolor[rgb]{ .996,  .91,  .514}0.9628 & \cellcolor[rgb]{ .631,  .816,  .498}0.8414 & \cellcolor[rgb]{ .722,  .843,  .502}0.9988 \\
    \midrule
    Power & \cellcolor[rgb]{ .973,  .412,  .42}0.8751 & \cellcolor[rgb]{ .973,  .412,  .42}0.0745 & \cellcolor[rgb]{ .973,  .412,  .42}0.8850 & \cellcolor[rgb]{ .973,  .431,  .424}0.6477 & \cellcolor[rgb]{ .992,  .788,  .49}0.9973 \\
    \midrule
    ShPower & \cellcolor[rgb]{ .996,  .863,  .506}0.9470 & \cellcolor[rgb]{ 1,  .871,  .51}0.0331 & \cellcolor[rgb]{ .996,  .859,  .506}0.9549 & \cellcolor[rgb]{ .992,  .824,  .498}0.7748 & \cellcolor[rgb]{ 1,  .922,  .518}0.9981 \\
    \midrule
    DGBD  & \cellcolor[rgb]{ .388,  .745,  .482}0.9689 & \cellcolor[rgb]{ .388,  .745,  .482}0.0196 & \cellcolor[rgb]{ .388,  .745,  .482}0.9736 & \cellcolor[rgb]{ .388,  .745,  .482}0.8641 & \cellcolor[rgb]{ .694,  .835,  .502}0.9988 \\
    \midrule
    ZML & \cellcolor[rgb]{ .765,  .855,  .506}0.9612 & \cellcolor[rgb]{ .867,  .882,  .51}0.0263 & \cellcolor[rgb]{ .718,  .843,  .502}0.9685 & \cellcolor[rgb]{ .737,  .847,  .506}0.8312 & \cellcolor[rgb]{ .388,  .745,  .482}0.9996 \\
    \midrule
    Poly & \cellcolor[rgb]{ .992,  .796,  .494}0.9369 & \cellcolor[rgb]{ .976,  .447,  .427}0.0716 & \cellcolor[rgb]{ 1,  .922,  .518}0.9641 & \cellcolor[rgb]{ .973,  .412,  .42}0.6406 & \cellcolor[rgb]{ .973,  .412,  .42}0.9949 \\
    \midrule
    Uni & \cellcolor[rgb]{ .561,  .796,  .494}0.9654 & \cellcolor[rgb]{ .902,  .89,  .51}0.0269 & \cellcolor[rgb]{ .408,  .753,  .486}0.9733 & \cellcolor[rgb]{ 1,  .922,  .518}0.8059 & \cellcolor[rgb]{ .996,  .918,  .514}0.9980 \\
    \bottomrule
    \end{tabular}%
    }
    \label{tab:R2_DJ}%
  \end{minipage}
  \hfill
  \begin{minipage}{0.45\textwidth}
    \caption{Summary statistics of the pooled RMSE values for the DJI dataset (lower values indicate better fit).}
  \resizebox{\textwidth}{!}{ 
   \begin{tabular}{|c|c|c|c|c|c|}
    \toprule
    RMSE  & $\mu$  & $\sigma$   & $Me$ & $\min$   & $\max$ \\
    \midrule
    Exp & \cellcolor[rgb]{ .776,  .855,  .502}0.0099 & \cellcolor[rgb]{ .388,  .745,  .482}0.0059 & \cellcolor[rgb]{ .655,  .82,  .494}0.0097 & \cellcolor[rgb]{ .973,  .412,  .42}0.0012 & \cellcolor[rgb]{ .388,  .745,  .482}0.0309 \\
    \midrule
    Power & \cellcolor[rgb]{ .973,  .412,  .42}0.0177 & \cellcolor[rgb]{ .992,  .753,  .486}0.0100 & \cellcolor[rgb]{ .973,  .412,  .42}0.0195 & \cellcolor[rgb]{ .98,  .514,  .439}0.0012 & \cellcolor[rgb]{ .996,  .82,  .498}0.0463 \\
    \midrule
    ShPower & \cellcolor[rgb]{ .996,  .792,  .494}0.0122 & \cellcolor[rgb]{ 1,  .863,  .506}0.0080 & \cellcolor[rgb]{ 1,  .863,  .51}0.0119 & \cellcolor[rgb]{ .988,  .698,  .478}0.0010 & \cellcolor[rgb]{ 1,  .878,  .51}0.0412 \\
    \midrule
    DGBD  & \cellcolor[rgb]{ .388,  .745,  .482}0.0092 & \cellcolor[rgb]{ .588,  .8,  .49}0.0062 & \cellcolor[rgb]{ .388,  .745,  .482}0.0088 & \cellcolor[rgb]{ .388,  .745,  .482}0.0004 & \cellcolor[rgb]{ .631,  .816,  .494}0.0335 \\
    \midrule
    ZML & \cellcolor[rgb]{ 1,  .922,  .518}0.0102 & \cellcolor[rgb]{ .604,  .804,  .494}0.0062 & \cellcolor[rgb]{ 1,  .922,  .518}0.0109 & \cellcolor[rgb]{ .871,  .882,  .51}0.0008 & \cellcolor[rgb]{ .663,  .824,  .498}0.0338 \\
    \midrule
    Poly & \cellcolor[rgb]{ .98,  .549,  .447}0.0157 & \cellcolor[rgb]{ .973,  .412,  .42}0.0164 & \cellcolor[rgb]{ 1,  .918,  .518}0.0110 & \cellcolor[rgb]{ .604,  .808,  .494}0.0006 & \cellcolor[rgb]{ .973,  .412,  .42}0.0815 \\
    \midrule
    Uni & \cellcolor[rgb]{ .902,  .89,  .51}0.0101 & \cellcolor[rgb]{ 1,  .922,  .518}0.0068 & \cellcolor[rgb]{ .659,  .824,  .498}0.0098 & \cellcolor[rgb]{ 1,  .922,  .518}0.0009 & \cellcolor[rgb]{ 1,  .922,  .518}0.0374 \\
    \bottomrule
    \end{tabular}%
    }
    \label{tab:RMSE_DJ}%
  \end{minipage}
\end{table}

\begin{table}[!htbp]
  \centering
  \begin{minipage}{0.45\textwidth}
    \caption{Summary statistics of the pooled MAE values for the DJI dataset (lower values indicate better fit).}
    \resizebox{\textwidth}{!}{ 
    \begin{tabular}{|c|c|c|c|c|c|}
    \toprule
    MAE   & $\mu$  & $\sigma$   & $Me$ & $\min$   & $\max$ \\
    \midrule
    Exp & \cellcolor[rgb]{ .561,  .792,  .49}0.0059 & \cellcolor[rgb]{ .388,  .745,  .482}0.0032 & \cellcolor[rgb]{ .627,  .812,  .494}0.0060 & \cellcolor[rgb]{ .973,  .412,  .42}0.0010 & \cellcolor[rgb]{ .388,  .745,  .482}0.0167 \\
    \midrule
    Power & \cellcolor[rgb]{ .973,  .412,  .42}0.0142 & \cellcolor[rgb]{ .984,  .604,  .459}0.0081 & \cellcolor[rgb]{ .973,  .412,  .42}0.0160 & \cellcolor[rgb]{ .988,  .671,  .471}0.0008 & \cellcolor[rgb]{ .992,  .71,  .478}0.0331 \\
    \midrule
    ShPower & \cellcolor[rgb]{ .996,  .804,  .498}0.0087 & \cellcolor[rgb]{ .996,  .804,  .498}0.0056 & \cellcolor[rgb]{ .996,  .824,  .502}0.0087 & \cellcolor[rgb]{ .996,  .812,  .498}0.0007 & \cellcolor[rgb]{ .996,  .8,  .494}0.0280 \\
    \midrule
    DGBD  & \cellcolor[rgb]{ .388,  .745,  .482}0.0055 & \cellcolor[rgb]{ .471,  .769,  .486}0.0033 & \cellcolor[rgb]{ .388,  .745,  .482}0.0054 & \cellcolor[rgb]{ .388,  .745,  .482}0.0003 & \cellcolor[rgb]{ .58,  .8,  .49}0.0180 \\
    \midrule
    ZML & \cellcolor[rgb]{ 1,  .922,  .518}0.0070 & \cellcolor[rgb]{ 1,  .922,  .518}0.0041 & \cellcolor[rgb]{ 1,  .89,  .514}0.0075 & \cellcolor[rgb]{ 1,  .922,  .518}0.0006 & \cellcolor[rgb]{ .784,  .859,  .502}0.0195 \\
    \midrule
    Poly & \cellcolor[rgb]{ .988,  .698,  .475}0.0102 & \cellcolor[rgb]{ .973,  .412,  .42}0.0104 & \cellcolor[rgb]{ 1,  .922,  .518}0.0069 & \cellcolor[rgb]{ .639,  .816,  .494}0.0004 & \cellcolor[rgb]{ .973,  .412,  .42}0.0499 \\
    \midrule
    Uni & \cellcolor[rgb]{ .635,  .816,  .494}0.0061 & \cellcolor[rgb]{ .784,  .859,  .502}0.0038 & \cellcolor[rgb]{ .722,  .839,  .498}0.0062 & \cellcolor[rgb]{ .992,  .918,  .514}0.0006 & \cellcolor[rgb]{ 1,  .922,  .518}0.0211 \\
    \bottomrule
    \end{tabular}%
    }
    \label{tab:MAE_DJ}%
  \end{minipage}
  \hfill
  \begin{minipage}{0.45\textwidth}
    \caption{Summary statistics of the pooled AIC values for the DJI dataset (lower values indicate better fit).}
  \resizebox{\textwidth}{!}{ 
   \begin{tabular}{|c|c|c|c|c|c|}
    \toprule
    AIC   & $\mu$  & $\sigma$   & $Me$ & $\min$   & $\max$ \\
    \midrule
    Exp & \cellcolor[rgb]{ 1,  .922,  .518}-279.95 & \cellcolor[rgb]{ .388,  .745,  .482}44.75 & \cellcolor[rgb]{ .612,  .808,  .494}-266.65 & \cellcolor[rgb]{ .973,  .412,  .42}-387.27 & \cellcolor[rgb]{ .388,  .745,  .482}-199.83 \\
    \midrule
    Power & \cellcolor[rgb]{ .973,  .412,  .42}-249.05 & \cellcolor[rgb]{ 1,  .914,  .518}52.79 & \cellcolor[rgb]{ .973,  .412,  .42}-226.50 & \cellcolor[rgb]{ .98,  .518,  .443}-390.60 & \cellcolor[rgb]{ .996,  .812,  .498}-176.28 \\
    \midrule
    ShPower & \cellcolor[rgb]{ .992,  .776,  .49}-271.04 & \cellcolor[rgb]{ .988,  .918,  .514}52.27 & \cellcolor[rgb]{ .996,  .843,  .506}-254.00 & \cellcolor[rgb]{ .988,  .694,  .475}-396.30 & \cellcolor[rgb]{ 1,  .878,  .51}-182.14 \\
    \midrule
    DGBD  & \cellcolor[rgb]{ .388,  .745,  .482}-286.65 & \cellcolor[rgb]{ .937,  .902,  .514}51.59 & \cellcolor[rgb]{ .388,  .745,  .482}-271.08 & \cellcolor[rgb]{ .388,  .745,  .482}-446.99 & \cellcolor[rgb]{ .686,  .831,  .498}-193.09 \\
    \midrule
    ZML & \cellcolor[rgb]{ .871,  .882,  .51}-281.34 & \cellcolor[rgb]{ 1,  .871,  .51}54.51 & \cellcolor[rgb]{ 1,  .922,  .518}-259.14 & \cellcolor[rgb]{ .882,  .886,  .51}-412.10 & \cellcolor[rgb]{ .663,  .824,  .498}-193.62 \\
    \midrule
    Poly & \cellcolor[rgb]{ .996,  .816,  .498}-273.30 & \cellcolor[rgb]{ .973,  .412,  .42}73.34 & \cellcolor[rgb]{ 1,  .902,  .518}-257.88 & \cellcolor[rgb]{ .659,  .824,  .498}-427.70 & \cellcolor[rgb]{ .973,  .412,  .42}-141.75 \\
    \midrule
    Uni & \cellcolor[rgb]{ .945,  .906,  .514}-280.53 & \cellcolor[rgb]{ 1,  .922,  .518}52.37 & \cellcolor[rgb]{ .749,  .847,  .502}-264.01 & \cellcolor[rgb]{ 1,  .922,  .518}-403.80 & \cellcolor[rgb]{ 1,  .922,  .518}-186.10 \\
    \bottomrule
    \end{tabular}%
    }
    \label{tab:AIC_DJ}%
  \end{minipage}
\end{table}

\begin{table}[!htbp]
  \centering
  \resizebox{0.8\textwidth}{!}{ 
    \begin{tabular}{|c|c|c|c|c|c|c|c|}
\cmidrule{2-8}    \multicolumn{1}{r|}{} & Exp & Power & ShPower & DGBD  & ZML & Poly & Uni \\
    \midrule
    Exp & -      & \cellcolor[rgb]{ .796,  .863,  .506}-0.0078 & \cellcolor[rgb]{ .796,  .863,  .506}-0.0023 & \cellcolor[rgb]{ .388,  .745,  .482}0.2683 & \cellcolor[rgb]{ .388,  .745,  .482}0.0070 & \cellcolor[rgb]{ .796,  .863,  .506}-0.0059 & \cellcolor[rgb]{ .388,  .745,  .482}-0.0003 \\
    \midrule
    Power & \cellcolor[rgb]{ .984,  .616,  .459}8.0318 & -      & \cellcolor[rgb]{ .985,  .641,  .464}4.9480 & \cellcolor[rgb]{ .984,  .616,  .459}8.2480 & \cellcolor[rgb]{ .984,  .616,  .459}7.7121 & \cellcolor[rgb]{ .973,  .412,  .42}0.5200 & \cellcolor[rgb]{ .984,  .616,  .459}7.1532 \\
    \midrule
    ShPower & \cellcolor[rgb]{ .997,  .866,  .507}3.0838 & \cellcolor[rgb]{ 1,  .922,  .518}-0.0050 & -      & \cellcolor[rgb]{ .997,  .866,  .507}3.3001 & \cellcolor[rgb]{ .997,  .867,  .507}2.7642 & \cellcolor[rgb]{ 1,  .92,  .518}-0.0036 & \cellcolor[rgb]{ .997,  .867,  .507}2.2053 \\
    \midrule
    DGBD & \cellcolor[rgb]{ .388,  .745,  .482}0.0521 & \cellcolor[rgb]{ .388,  .745,  .482}-0.0085 & \cellcolor[rgb]{ .388,  .745,  .482}-0.0029 & -      & \cellcolor[rgb]{ .401,  .749,  .483}0.0419 & \cellcolor[rgb]{ .388,  .745,  .482}-0.0065 & \cellcolor[rgb]{ .388,  .745,  .482}-0.0009 \\
    \midrule
    ZML & \cellcolor[rgb]{ .475,  .77,  .487}0.3267 & \cellcolor[rgb]{ 1,  .922,  .518}-0.0075 & \cellcolor[rgb]{ 1,  .922,  .518}-0.0019 & \cellcolor[rgb]{ .486,  .773,  .488}0.5778 & -      & \cellcolor[rgb]{ 1,  .922,  .518}-0.0055 & \cellcolor[rgb]{ .404,  .75,  .483}0.0282 \\
    \midrule
    Poly & \cellcolor[rgb]{ .973,  .412,  .42}12.0540 & \cellcolor[rgb]{ .973,  .412,  .42}4.5421 & \cellcolor[rgb]{ .973,  .412,  .42}8.9701 & \cellcolor[rgb]{ .973,  .412,  .42}12.2702 & \cellcolor[rgb]{ .973,  .412,  .42}11.7343 & -      & \cellcolor[rgb]{ .973,  .412,  .42}11.1754 \\
    \midrule
    Uni & \cellcolor[rgb]{ .65,  .821,  .497}0.8786 & \cellcolor[rgb]{ .974,  .914,  .516}-0.0075 & \cellcolor[rgb]{ .974,  .914,  .516}-0.0020 & \cellcolor[rgb]{ .65,  .821,  .497}1.0948 & \cellcolor[rgb]{ .601,  .807,  .495}0.5871 & \cellcolor[rgb]{ .974,  .914,  .516}-0.0056 & -      \\
    \bottomrule
    \end{tabular}%
    }
    \caption{Pairwise signed dominance indices for the DJI dataset.
Each entry \((i,j)\) reports
\(\mathcal{D}(X^{(i)},X^{(j)})\).}
  \label{tab:pairwise_dominance_DJ}%
\end{table}%


\begin{table}[!htbp]
  \centering
  \begin{minipage}{0.45\textwidth}
    \caption{Summary statistics of the pooled $R^2$ values for the STOXX50E dataset (higher values indicate better fit).}
    \resizebox{\textwidth}{!}{ 
    \begin{tabular}{|c|c|c|c|c|c|}
    \toprule
    $R^2$    & $\mu$  & $\sigma$   & $Me$ & $\min$   & $\max$ \\
      \midrule
    Exp & \cellcolor[rgb]{ 1,  .922,  .518}0.9452 & \cellcolor[rgb]{ .996,  .804,  .498}0.0525 & \cellcolor[rgb]{ .388,  .745,  .482}0.9650 & \cellcolor[rgb]{ .992,  .792,  .49}0.6245 & \cellcolor[rgb]{ .416,  .753,  .486}0.9989 \\
    \midrule
    Power & \cellcolor[rgb]{ .973,  .412,  .42}0.8732 & \cellcolor[rgb]{ .988,  .655,  .467}0.0689 & \cellcolor[rgb]{ .973,  .412,  .42}0.8820 & \cellcolor[rgb]{ 1,  .922,  .518}0.6612 & \cellcolor[rgb]{ .973,  .412,  .42}0.9918 \\
    \midrule
    ShPower & \cellcolor[rgb]{ .992,  .812,  .494}0.9302 & \cellcolor[rgb]{ 1,  .922,  .518}0.0396 & \cellcolor[rgb]{ .992,  .78,  .49}0.9376 & \cellcolor[rgb]{ .663,  .824,  .498}0.7635 & \cellcolor[rgb]{ .973,  .424,  .42}0.9919 \\
    \midrule
    DGBD  & \cellcolor[rgb]{ .478,  .773,  .49}0.9581 & \cellcolor[rgb]{ .643,  .82,  .494}0.0297 & \cellcolor[rgb]{ .396,  .749,  .486}0.9649 & \cellcolor[rgb]{ .608,  .808,  .498}0.7797 & \cellcolor[rgb]{ 1,  .922,  .518}0.9962 \\
    \midrule
    ZML & \cellcolor[rgb]{ .388,  .745,  .482}0.9602 & \cellcolor[rgb]{ .388,  .745,  .482}0.0225 & \cellcolor[rgb]{ .463,  .769,  .49}0.9643 & \cellcolor[rgb]{ .388,  .745,  .482}0.8449 & \cellcolor[rgb]{ .388,  .745,  .482}0.9990 \\
    \midrule
    Poly & \cellcolor[rgb]{ .98,  .565,  .447}0.8949 & \cellcolor[rgb]{ .973,  .412,  .42}0.0948 & \cellcolor[rgb]{ .984,  .69,  .471}0.9244 & \cellcolor[rgb]{ .973,  .412,  .42}0.5123 & \cellcolor[rgb]{ .992,  .816,  .494}0.9953 \\
    \midrule
    Uni & \cellcolor[rgb]{ .792,  .863,  .506}0.9503 & \cellcolor[rgb]{ .929,  .902,  .514}0.0377 & \cellcolor[rgb]{ 1,  .922,  .518}0.9586 & \cellcolor[rgb]{ .996,  .894,  .51}0.6542 & \cellcolor[rgb]{ .957,  .91,  .518}0.9964 \\
    \bottomrule
    \end{tabular}%
    }
    \label{tab:R2_EX}%
  \end{minipage}
  \hfill
  \begin{minipage}{0.45\textwidth}
    \caption{Summary statistics of the pooled RMSE values for the STOXX50E dataset (lower values indicate better fit).}
  \resizebox{\textwidth}{!}{ 
   \begin{tabular}{|c|c|c|c|c|c|}
    \toprule
    RMSE  & $\mu$  & $\sigma$   & $Me$ & $\min$   & $\max$ \\
     \midrule
    Exp & \cellcolor[rgb]{ .388,  .745,  .482}0.0077 & \cellcolor[rgb]{ .388,  .745,  .482}0.0044 & \cellcolor[rgb]{ .388,  .745,  .482}0.0075 & \cellcolor[rgb]{ .973,  .412,  .42}0.0008 & \cellcolor[rgb]{ .388,  .745,  .482}0.0320 \\
    \midrule
    Power & \cellcolor[rgb]{ .976,  .431,  .424}0.0150 & \cellcolor[rgb]{ .996,  .788,  .494}0.0087 & \cellcolor[rgb]{ .973,  .412,  .42}0.0169 & \cellcolor[rgb]{ .98,  .541,  .447}0.0008 & \cellcolor[rgb]{ 1,  .886,  .514}0.0425 \\
    \midrule
    ShPower & \cellcolor[rgb]{ .992,  .741,  .482}0.0114 & \cellcolor[rgb]{ 1,  .875,  .51}0.0075 & \cellcolor[rgb]{ .992,  .757,  .486}0.0117 & \cellcolor[rgb]{ .992,  .769,  .49}0.0007 & \cellcolor[rgb]{ 1,  .902,  .518}0.0414 \\
    \midrule
    DGBD  & \cellcolor[rgb]{ .933,  .902,  .514}0.0092 & \cellcolor[rgb]{ 1,  .922,  .518}0.0068 & \cellcolor[rgb]{ .745,  .847,  .502}0.0085 & \cellcolor[rgb]{ .42,  .753,  .482}0.0004 & \cellcolor[rgb]{ .976,  .914,  .514}0.0400 \\
    \midrule
    ZML & \cellcolor[rgb]{ .561,  .792,  .49}0.0082 & \cellcolor[rgb]{ .576,  .796,  .49}0.0051 & \cellcolor[rgb]{ .776,  .855,  .502}0.0086 & \cellcolor[rgb]{ 1,  .922,  .518}0.0006 & \cellcolor[rgb]{ .545,  .788,  .49}0.0341 \\
    \midrule
    Poly & \cellcolor[rgb]{ .973,  .412,  .42}0.0152 & \cellcolor[rgb]{ .973,  .412,  .42}0.0138 & \cellcolor[rgb]{ .992,  .718,  .478}0.0123 & \cellcolor[rgb]{ .388,  .745,  .482}0.0003 & \cellcolor[rgb]{ .973,  .412,  .42}0.0708 \\
    \midrule
    Uni & \cellcolor[rgb]{ 1,  .922,  .518}0.0093 & \cellcolor[rgb]{ .875,  .882,  .51}0.0063 & \cellcolor[rgb]{ 1,  .922,  .518}0.0092 & \cellcolor[rgb]{ .784,  .859,  .502}0.0005 & \cellcolor[rgb]{ 1,  .922,  .518}0.0402 \\
    \bottomrule
    \end{tabular}%
    }
    \label{tab:RMSE_EX}%
  \end{minipage}
\end{table}

\begin{table}[!htbp]
  \centering
  \begin{minipage}{0.45\textwidth}
    \caption{Summary statistics of the pooled MAE values for the STOXX50E dataset (lower values indicate better fit).}
    \resizebox{\textwidth}{!}{ 
    \begin{tabular}{|c|c|c|c|c|c|}
    \toprule
    MAE   & $\mu$  & $\sigma$   & $Me$ & $\min$   & $\max$ \\
    \midrule
    Exp & \cellcolor[rgb]{ .388,  .745,  .482}0.0039 & \cellcolor[rgb]{ .388,  .745,  .482}0.0018 & \cellcolor[rgb]{ .388,  .745,  .482}0.0039 & \cellcolor[rgb]{ .973,  .412,  .42}0.0006 & \cellcolor[rgb]{ .388,  .745,  .482}0.0104 \\
    \midrule
    Power & \cellcolor[rgb]{ .973,  .412,  .42}0.0113 & \cellcolor[rgb]{ .988,  .667,  .471}0.0062 & \cellcolor[rgb]{ .973,  .412,  .42}0.0135 & \cellcolor[rgb]{ .976,  .459,  .431}0.0006 & \cellcolor[rgb]{ 1,  .878,  .51}0.0222 \\
    \midrule
    ShPower & \cellcolor[rgb]{ .988,  .702,  .478}0.0079 & \cellcolor[rgb]{ .996,  .796,  .494}0.0047 & \cellcolor[rgb]{ .992,  .729,  .482}0.0086 & \cellcolor[rgb]{ 1,  .922,  .518}0.0005 & \cellcolor[rgb]{ 1,  .922,  .518}0.0202 \\
    \midrule
    DGBD  & \cellcolor[rgb]{ .847,  .875,  .506}0.0050 & \cellcolor[rgb]{ 1,  .922,  .518}0.0033 & \cellcolor[rgb]{ .737,  .843,  .502}0.0049 & \cellcolor[rgb]{ .388,  .745,  .482}0.0003 & \cellcolor[rgb]{ .745,  .847,  .502}0.0162 \\
    \midrule
    ZML & \cellcolor[rgb]{ .769,  .855,  .502}0.0048 & \cellcolor[rgb]{ .733,  .843,  .502}0.0026 & \cellcolor[rgb]{ .902,  .894,  .51}0.0054 & \cellcolor[rgb]{ .996,  .824,  .502}0.0005 & \cellcolor[rgb]{ .494,  .776,  .486}0.0121 \\
    \midrule
    Poly & \cellcolor[rgb]{ .98,  .498,  .439}0.0103 & \cellcolor[rgb]{ .973,  .412,  .42}0.0091 & \cellcolor[rgb]{ .992,  .749,  .486}0.0083 & \cellcolor[rgb]{ .396,  .745,  .482}0.0003 & \cellcolor[rgb]{ .973,  .412,  .42}0.0426 \\
    \midrule
    Uni & \cellcolor[rgb]{ 1,  .922,  .518}0.0054 & \cellcolor[rgb]{ .941,  .902,  .514}0.0031 & \cellcolor[rgb]{ 1,  .922,  .518}0.0057 & \cellcolor[rgb]{ .788,  .859,  .502}0.0004 & \cellcolor[rgb]{ 1,  .878,  .51}0.0223 \\
    \bottomrule
    \end{tabular}%
    }
    \label{tab:MAE_EX}%
  \end{minipage}
  \hfill
  \begin{minipage}{0.45\textwidth}
    \caption{Summary statistics of the pooled AIC values for the STOXX50E dataset (lower values indicate better fit).}
  \resizebox{\textwidth}{!}{ 
   \begin{tabular}{|c|c|c|c|c|c|}
    \toprule
    AIC   & $\mu$  & $\sigma$   & $Me$ & $\min$   & $\max$ \\
    \midrule
    Exp & \cellcolor[rgb]{ .627,  .812,  .494}-474.54 & \cellcolor[rgb]{ .388,  .745,  .482}66.26 & \cellcolor[rgb]{ .388,  .745,  .482}-457.83 & \cellcolor[rgb]{ .973,  .412,  .42}-666.17 & \cellcolor[rgb]{ .388,  .745,  .482}-321.72 \\
    \midrule
    Power & \cellcolor[rgb]{ .973,  .412,  .42}-426.73 & \cellcolor[rgb]{ 1,  .898,  .514}98.56 & \cellcolor[rgb]{ .973,  .412,  .42}-381.54 & \cellcolor[rgb]{ .98,  .533,  .443}-671.35 & \cellcolor[rgb]{ 1,  .902,  .514}-294.99 \\
    \midrule
    ShPower & \cellcolor[rgb]{ .992,  .745,  .486}-452.64 & \cellcolor[rgb]{ 1,  .922,  .518}97.56 & \cellcolor[rgb]{ .992,  .729,  .482}-415.47 & \cellcolor[rgb]{ .992,  .749,  .486}-680.45 & \cellcolor[rgb]{ 1,  .914,  .518}-296.35 \\
    \midrule
    DGBD  & \cellcolor[rgb]{ .388,  .745,  .482}-479.80 & \cellcolor[rgb]{ .988,  .671,  .471}107.02 & \cellcolor[rgb]{ .765,  .851,  .502}-444.51 & \cellcolor[rgb]{ .424,  .753,  .482}-742.19 & \cellcolor[rgb]{ .961,  .91,  .514}-298.72 \\
    \midrule
    ZML & \cellcolor[rgb]{ .435,  .757,  .482}-478.70 & \cellcolor[rgb]{ .792,  .859,  .502}87.06 & \cellcolor[rgb]{ .769,  .851,  .502}-444.48 & \cellcolor[rgb]{ 1,  .922,  .518}-687.90 & \cellcolor[rgb]{ .565,  .796,  .49}-314.60 \\
    \midrule
    Poly & \cellcolor[rgb]{ .984,  .62,  .463}-442.83 & \cellcolor[rgb]{ .973,  .412,  .42}116.58 & \cellcolor[rgb]{ .988,  .675,  .471}-409.53 & \cellcolor[rgb]{ .388,  .745,  .482}-745.51 & \cellcolor[rgb]{ .973,  .412,  .42}-245.04 \\
    \midrule
    Uni & \cellcolor[rgb]{ 1,  .922,  .518}-466.37 & \cellcolor[rgb]{ .843,  .875,  .506}89.54 & \cellcolor[rgb]{ 1,  .922,  .518}-436.37 & \cellcolor[rgb]{ .839,  .875,  .506}-702.67 & \cellcolor[rgb]{ 1,  .922,  .518}-297.30 \\
    \bottomrule
    \end{tabular}%
    }
    \label{tab:AIC_EX}%
  \end{minipage}
\end{table}

\begin{table}[!htbp]
  \centering
  \resizebox{0.8\textwidth}{!}{ 
    \begin{tabular}{|c|c|c|c|c|c|c|c|}
\cmidrule{2-8}    \multicolumn{1}{r|}{} & Exp & Power & ShPower & DGBD  & ZML & Poly & Uni \\
    \midrule
    Exp & -      & \cellcolor[rgb]{ .388,  .745,  .482}-0.0073 & \cellcolor[rgb]{ .388,  .745,  .482}-0.0037 & \cellcolor[rgb]{ .388,  .745,  .482}-0.0015 & \cellcolor[rgb]{ .388,  .745,  .482}-0.0005 & \cellcolor[rgb]{ .388,  .745,  .482}-0.0075 & \cellcolor[rgb]{ .388,  .745,  .482}-0.0015 \\
    \midrule
    Power & \cellcolor[rgb]{ .988,  .693,  .474}7.5890 & -      & \cellcolor[rgb]{ .99,  .728,  .481}2.9066 & \cellcolor[rgb]{ .988,  .693,  .474}4.8290 & \cellcolor[rgb]{ .988,  .693,  .474}6.8410 & \cellcolor[rgb]{ .973,  .412,  .42}-0.0002 & \cellcolor[rgb]{ .988,  .693,  .474}5.1572 \\
    \midrule
    ShPower & \cellcolor[rgb]{ .997,  .865,  .507}4.6824 & \cellcolor[rgb]{ 1,  .922,  .518}-0.0010 & -      & \cellcolor[rgb]{ .997,  .865,  .507}1.9223 & \cellcolor[rgb]{ .997,  .865,  .507}3.9344 & \cellcolor[rgb]{ .99,  .725,  .48}-0.0038 & \cellcolor[rgb]{ .997,  .865,  .507}2.2505 \\
    \midrule
    DGBD & \cellcolor[rgb]{ .802,  .865,  .506}2.7600 & \cellcolor[rgb]{ 1,  .922,  .518}-0.0025 & \cellcolor[rgb]{ 1,  .922,  .518}-0.0014 & -      & \cellcolor[rgb]{ .802,  .865,  .506}2.0120 & \cellcolor[rgb]{ .982,  .917,  .517}-0.0060 & \cellcolor[rgb]{ .548,  .791,  .491}0.3368 \\
    \midrule
    ZML & \cellcolor[rgb]{ .388,  .745,  .482}0.7480 & \cellcolor[rgb]{ .46,  .766,  .486}-0.0068 & \cellcolor[rgb]{ .532,  .787,  .49}-0.0033 & \cellcolor[rgb]{ .388,  .745,  .482}-0.0010 & -      & \cellcolor[rgb]{ .576,  .799,  .493}-0.0070 & \cellcolor[rgb]{ .388,  .745,  .482}-0.0011 \\
    \midrule
    Poly & \cellcolor[rgb]{ .973,  .412,  .42}12.3325 & \cellcolor[rgb]{ .973,  .412,  .42}4.7435 & \cellcolor[rgb]{ .973,  .412,  .42}7.6502 & \cellcolor[rgb]{ .973,  .412,  .42}9.5725 & \cellcolor[rgb]{ .973,  .412,  .42}11.5845 & -      & \cellcolor[rgb]{ .973,  .412,  .42}9.9007 \\
    \midrule
    Uni & \cellcolor[rgb]{ .735,  .845,  .502}2.4318 & \cellcolor[rgb]{ .864,  .883,  .51}-0.0043 & \cellcolor[rgb]{ .871,  .885,  .51}-0.0022 & \cellcolor[rgb]{ .394,  .747,  .482}0.0086 & \cellcolor[rgb]{ .735,  .845,  .502}1.6838 & \cellcolor[rgb]{ 1,  .918,  .517}-0.0059 & -      \\
    \bottomrule
    \end{tabular}%
    }
    \caption{Pairwise signed dominance indices for the STOXX50E dataset.
Each entry \((i,j)\) reports
\(\mathcal{D}(X^{(i)},X^{(j)})\).}
  \label{tab:pairwise_dominance_EX}%
\end{table}%


\begin{table}[!htbp]
  \centering
  \begin{minipage}{0.45\textwidth}
    \caption{Summary statistics of the pooled $R^2$ values for the NDX dataset (higher values indicate better fit).}
    \resizebox{\textwidth}{!}{ 
    \begin{tabular}{|c|c|c|c|c|c|}
    \toprule
    $R^2$    & $\mu$  & $\sigma$   & $Me$ & $\min$   & $\max$ \\
        \midrule
    Exp & \cellcolor[rgb]{ 1,  .922,  .518}0.9382 & \cellcolor[rgb]{ .992,  .761,  .486}0.0679 & \cellcolor[rgb]{ .749,  .851,  .506}0.9652 & \cellcolor[rgb]{ .984,  .659,  .467}0.6228 & \cellcolor[rgb]{ .388,  .745,  .482}0.9983 \\
    \midrule
    Power & \cellcolor[rgb]{ .984,  .663,  .467}0.8852 & \cellcolor[rgb]{ .996,  .796,  .494}0.0607 & \cellcolor[rgb]{ .976,  .545,  .443}0.8895 & \cellcolor[rgb]{ .992,  .788,  .49}0.7075 & \cellcolor[rgb]{ .973,  .412,  .42}0.9892 \\
    \midrule
    ShPower & \cellcolor[rgb]{ .996,  .91,  .514}0.9365 & \cellcolor[rgb]{ 1,  .922,  .518}0.0375 & \cellcolor[rgb]{ .992,  .824,  .498}0.9432 & \cellcolor[rgb]{ 1,  .922,  .518}0.7945 & \cellcolor[rgb]{ .996,  .847,  .502}0.9943 \\
    \midrule
    DGBD  & \cellcolor[rgb]{ .647,  .82,  .498}0.9546 & \cellcolor[rgb]{ .8,  .863,  .506}0.0318 & \cellcolor[rgb]{ .91,  .898,  .514}0.9630 & \cellcolor[rgb]{ .91,  .898,  .514}0.8058 & \cellcolor[rgb]{ 1,  .922,  .518}0.9952 \\
    \midrule
    ZML & \cellcolor[rgb]{ .388,  .745,  .482}0.9664 & \cellcolor[rgb]{ .388,  .745,  .482}0.0197 & \cellcolor[rgb]{ .388,  .745,  .482}0.9701 & \cellcolor[rgb]{ .388,  .745,  .482}0.8686 & \cellcolor[rgb]{ .42,  .757,  .486}0.9981 \\
    \midrule
    Poly & \cellcolor[rgb]{ .973,  .412,  .42}0.8336 & \cellcolor[rgb]{ .973,  .412,  .42}0.1316 & \cellcolor[rgb]{ .973,  .412,  .42}0.8638 & \cellcolor[rgb]{ .973,  .412,  .42}0.4577 & \cellcolor[rgb]{ .996,  .882,  .51}0.9948 \\
    \midrule
    Uni & \cellcolor[rgb]{ .604,  .808,  .498}0.9565 & \cellcolor[rgb]{ .667,  .824,  .498}0.0279 & \cellcolor[rgb]{ 1,  .922,  .518}0.9618 & \cellcolor[rgb]{ .851,  .878,  .51}0.8129 & \cellcolor[rgb]{ .925,  .902,  .514}0.9956 \\
    \bottomrule
    \end{tabular}%
    }
    \label{tab:R2_NS}%
  \end{minipage}
  \hfill
  \begin{minipage}{0.45\textwidth}
    \caption{Summary statistics of the pooled RMSE values for the NDX dataset (lower values indicate better fit).}
  \resizebox{\textwidth}{!}{ 
   \begin{tabular}{|c|c|c|c|c|c|}
   \toprule
    RMSE   & $\mu$  & $\sigma$   & $Me$ & $\min$   & $\max$ \\
    \midrule
    Exp & \cellcolor[rgb]{ .506,  .776,  .486}0.0060 & \cellcolor[rgb]{ .396,  .745,  .482}0.0037 & \cellcolor[rgb]{ .388,  .745,  .482}0.0054 & \cellcolor[rgb]{ .973,  .412,  .42}0.0007 & \cellcolor[rgb]{ .388,  .745,  .482}0.0222 \\
    \midrule
    Power & \cellcolor[rgb]{ .988,  .678,  .475}0.0107 & \cellcolor[rgb]{ .996,  .835,  .502}0.0064 & \cellcolor[rgb]{ .976,  .471,  .431}0.0117 & \cellcolor[rgb]{ .98,  .553,  .447}0.0006 & \cellcolor[rgb]{ 1,  .89,  .514}0.0322 \\
    \midrule
    ShPower & \cellcolor[rgb]{ .996,  .843,  .502}0.0084 & \cellcolor[rgb]{ 1,  .894,  .514}0.0057 & \cellcolor[rgb]{ .992,  .776,  .49}0.0083 & \cellcolor[rgb]{ 1,  .922,  .518}0.0004 & \cellcolor[rgb]{ 1,  .914,  .518}0.0313 \\
    \midrule
    DGBD  & \cellcolor[rgb]{ 1,  .922,  .518}0.0072 & \cellcolor[rgb]{ 1,  .922,  .518}0.0053 & \cellcolor[rgb]{ 1,  .922,  .518}0.0067 & \cellcolor[rgb]{ .388,  .745,  .482}0.0002 & \cellcolor[rgb]{ 1,  .922,  .518}0.0309 \\
    \midrule
    ZML & \cellcolor[rgb]{ .388,  .745,  .482}0.0057 & \cellcolor[rgb]{ .388,  .745,  .482}0.0037 & \cellcolor[rgb]{ .545,  .788,  .49}0.0057 & \cellcolor[rgb]{ .996,  .816,  .498}0.0005 & \cellcolor[rgb]{ .514,  .78,  .486}0.0240 \\
    \midrule
    Poly & \cellcolor[rgb]{ .973,  .412,  .42}0.0146 & \cellcolor[rgb]{ .973,  .412,  .42}0.0120 & \cellcolor[rgb]{ .973,  .412,  .42}0.0123 & \cellcolor[rgb]{ .675,  .827,  .498}0.0003 & \cellcolor[rgb]{ .973,  .412,  .42}0.0520 \\
    \midrule
    Uni & \cellcolor[rgb]{ .761,  .851,  .502}0.0066 & \cellcolor[rgb]{ .69,  .831,  .498}0.0045 & \cellcolor[rgb]{ .898,  .89,  .51}0.0065 & \cellcolor[rgb]{ .749,  .847,  .502}0.0004 & \cellcolor[rgb]{ .914,  .894,  .51}0.0297 \\
    \bottomrule
    \end{tabular}%
    }
    \label{tab:RMSE_NS}%
  \end{minipage}
\end{table}
\begin{table}[!htbp]
  \centering
  \begin{minipage}{0.45\textwidth}
    \caption{Summary statistics of the pooled MAE values for the NDX dataset (lower values indicate better fit).}
    \resizebox{\textwidth}{!}{ 
    \begin{tabular}{|c|c|c|c|c|c|}
    \toprule
    MAE   & $\mu$  & $\sigma$   & $Me$ & $\min$   & $\max$ \\
    \midrule
    Exp & \cellcolor[rgb]{ .388,  .745,  .482}0.0026 & \cellcolor[rgb]{ .388,  .745,  .482}0.0013 & \cellcolor[rgb]{ .388,  .745,  .482}0.0026 & \cellcolor[rgb]{ .973,  .412,  .42}0.0005 & \cellcolor[rgb]{ .388,  .745,  .482}0.0077 \\
    \midrule
    Power & \cellcolor[rgb]{ .98,  .553,  .447}0.0077 & \cellcolor[rgb]{ .992,  .722,  .482}0.0043 & \cellcolor[rgb]{ .973,  .412,  .42}0.0091 & \cellcolor[rgb]{ .98,  .541,  .447}0.0005 & \cellcolor[rgb]{ .996,  .82,  .498}0.0159 \\
    \midrule
    ShPower & \cellcolor[rgb]{ .992,  .753,  .486}0.0056 & \cellcolor[rgb]{ .996,  .816,  .498}0.0034 & \cellcolor[rgb]{ .992,  .722,  .482}0.0060 & \cellcolor[rgb]{ 1,  .922,  .518}0.0003 & \cellcolor[rgb]{ 1,  .863,  .51}0.0144 \\
    \midrule
    DGBD  & \cellcolor[rgb]{ 1,  .922,  .518}0.0037 & \cellcolor[rgb]{ 1,  .922,  .518}0.0024 & \cellcolor[rgb]{ 1,  .922,  .518}0.0039 & \cellcolor[rgb]{ .388,  .745,  .482}0.0001 & \cellcolor[rgb]{ .925,  .898,  .51}0.0117 \\
    \midrule
    ZML & \cellcolor[rgb]{ .667,  .824,  .498}0.0031 & \cellcolor[rgb]{ .671,  .824,  .498}0.0018 & \cellcolor[rgb]{ .753,  .847,  .502}0.0033 & \cellcolor[rgb]{ .996,  .784,  .494}0.0004 & \cellcolor[rgb]{ .412,  .749,  .482}0.0078 \\
    \midrule
    Poly & \cellcolor[rgb]{ .973,  .412,  .42}0.0092 & \cellcolor[rgb]{ .973,  .412,  .42}0.0073 & \cellcolor[rgb]{ .98,  .525,  .443}0.0080 & \cellcolor[rgb]{ .506,  .776,  .486}0.0002 & \cellcolor[rgb]{ .973,  .412,  .42}0.0302 \\
    \midrule
    Uni & \cellcolor[rgb]{ .957,  .906,  .514}0.0037 & \cellcolor[rgb]{ .824,  .871,  .506}0.0021 & \cellcolor[rgb]{ .984,  .914,  .514}0.0038 & \cellcolor[rgb]{ .827,  .871,  .506}0.0003 & \cellcolor[rgb]{ 1,  .922,  .518}0.0123 \\
    \bottomrule
    \end{tabular}%
    }
    \label{tab:MAE_NS}%
  \end{minipage}
  \hfill
  \begin{minipage}{0.45\textwidth}
    \caption{Summary statistics of the pooled AIC values for the NDX dataset (lower values indicate better fit).}
  \resizebox{\textwidth}{!}{ 
   \begin{tabular}{|c|c|c|c|c|c|}
    \toprule
    AIC   & $\mu$  & $\sigma$   & $Me$ & $\min$   & $\max$ \\
    \midrule
    Exp & \cellcolor[rgb]{ 1,  .922,  .518}-767.11 & \cellcolor[rgb]{ .388,  .745,  .482}103.70 & \cellcolor[rgb]{ .388,  .745,  .482}-750.56 & \cellcolor[rgb]{ .973,  .412,  .42}-1047.43 & \cellcolor[rgb]{ .388,  .745,  .482}-546.64 \\
    \midrule
    Power & \cellcolor[rgb]{ .98,  .494,  .435}-702.63 & \cellcolor[rgb]{ 1,  .922,  .518}147.77 & \cellcolor[rgb]{ .976,  .467,  .431}-638.46 & \cellcolor[rgb]{ .98,  .533,  .443}-1061.99 & \cellcolor[rgb]{ 1,  .89,  .514}-492.64 \\
    \midrule
    ShPower & \cellcolor[rgb]{ .992,  .773,  .49}-744.70 & \cellcolor[rgb]{ .996,  .827,  .502}156.44 & \cellcolor[rgb]{ .992,  .749,  .486}-686.66 & \cellcolor[rgb]{ 1,  .922,  .518}-1108.76 & \cellcolor[rgb]{ 1,  .914,  .518}-496.11 \\
    \midrule
    DGBD  & \cellcolor[rgb]{ .855,  .878,  .506}-772.08 & \cellcolor[rgb]{ .992,  .733,  .482}164.76 & \cellcolor[rgb]{ 1,  .922,  .518}-717.09 & \cellcolor[rgb]{ .388,  .745,  .482}-1207.47 & \cellcolor[rgb]{ 1,  .922,  .518}-497.65 \\
    \midrule
    ZML & \cellcolor[rgb]{ .388,  .745,  .482}-788.33 & \cellcolor[rgb]{ .824,  .871,  .506}135.23 & \cellcolor[rgb]{ .569,  .796,  .49}-740.58 & \cellcolor[rgb]{ .996,  .792,  .494}-1092.88 & \cellcolor[rgb]{ .541,  .788,  .49}-534.18 \\
    \midrule
    Poly & \cellcolor[rgb]{ .973,  .412,  .42}-690.68 & \cellcolor[rgb]{ .973,  .412,  .42}193.03 & \cellcolor[rgb]{ .973,  .412,  .42}-629.07 & \cellcolor[rgb]{ .733,  .843,  .502}-1151.70 & \cellcolor[rgb]{ .973,  .412,  .42}-421.72 \\
    \midrule
    Uni & \cellcolor[rgb]{ .969,  .91,  .514}-768.13 & \cellcolor[rgb]{ .898,  .89,  .51}140.44 & \cellcolor[rgb]{ .937,  .902,  .514}-720.48 & \cellcolor[rgb]{ .812,  .867,  .506}-1139.11 & \cellcolor[rgb]{ .949,  .906,  .514}-501.56 \\
    \bottomrule
    \end{tabular}%
    }
    \label{tab:AIC_NS}%
  \end{minipage}\end{table}
\begin{table}[!htbp]
  \centering
   \resizebox{0.8\textwidth}{!}{ 
    \begin{tabular}{|c|c|c|c|c|c|c|c|}
\cmidrule{2-8}    \multicolumn{1}{r|}{} & Exp & Power & ShPower & DGBD  & ZML & Poly & Uni \\
    \midrule
    Exp & -      & \cellcolor[rgb]{ .461,  .766,  .486}-0.0047 & \cellcolor[rgb]{ .497,  .777,  .488}-0.0024 & \cellcolor[rgb]{ .388,  .745,  .482}-0.0012 & \cellcolor[rgb]{ .388,  .745,  .482}0.3494 & \cellcolor[rgb]{ .535,  .788,  .491}-0.0086 & \cellcolor[rgb]{ .388,  .745,  .482}-0.0007 \\
    \midrule
    Power & \cellcolor[rgb]{ .993,  .789,  .492}4.9643 & -      & \cellcolor[rgb]{ .994,  .811,  .497}1.9700 & \cellcolor[rgb]{ .993,  .789,  .492}3.0203 & \cellcolor[rgb]{ .993,  .789,  .492}5.3055 & \cellcolor[rgb]{ .973,  .412,  .42}-0.0039 & \cellcolor[rgb]{ .993,  .789,  .492}3.9698 \\
    \midrule
    ShPower & \cellcolor[rgb]{ .999,  .894,  .513}2.9944 & \cellcolor[rgb]{ 1,  .922,  .518}-0.0010 & -      & \cellcolor[rgb]{ .999,  .894,  .513}1.0503 & \cellcolor[rgb]{ .999,  .894,  .513}3.3356 & \cellcolor[rgb]{ .99,  .728,  .481}-0.0062 & \cellcolor[rgb]{ .999,  .894,  .513}1.9998 \\
    \midrule
    DGBD & \cellcolor[rgb]{ .869,  .884,  .51}1.9441 & \cellcolor[rgb]{ 1,  .922,  .518}-0.0013 & \cellcolor[rgb]{ 1,  .922,  .518}-0.0003 & -      & \cellcolor[rgb]{ .869,  .884,  .51}2.2853 & \cellcolor[rgb]{ .998,  .887,  .511}-0.0074 & \cellcolor[rgb]{ .782,  .859,  .505}0.9495 \\
    \midrule
    ZML & \cellcolor[rgb]{ .388,  .745,  .482}0.0081 & \cellcolor[rgb]{ .388,  .745,  .482}-0.0050 & \cellcolor[rgb]{ .388,  .745,  .482}-0.0027 & \cellcolor[rgb]{ .388,  .745,  .482}-0.0015 & -      & \cellcolor[rgb]{ .388,  .745,  .482}-0.0089 & \cellcolor[rgb]{ .388,  .745,  .482}-0.0010 \\
    \midrule
    Poly & \cellcolor[rgb]{ .973,  .412,  .42}12.0397 & \cellcolor[rgb]{ .973,  .412,  .42}7.0754 & \cellcolor[rgb]{ .973,  .412,  .42}9.0454 & \cellcolor[rgb]{ .973,  .412,  .42}10.0957 & \cellcolor[rgb]{ .973,  .412,  .42}12.3810 & -      & \cellcolor[rgb]{ .973,  .412,  .42}11.0452 \\
    \midrule
    Uni & \cellcolor[rgb]{ .633,  .816,  .496}0.9946 & \cellcolor[rgb]{ .693,  .833,  .5}-0.0038 & \cellcolor[rgb]{ .743,  .848,  .503}-0.0017 & \cellcolor[rgb]{ .389,  .745,  .482}-0.0005 & \cellcolor[rgb]{ .633,  .816,  .496}1.3358 & \cellcolor[rgb]{ .868,  .884,  .51}-0.0079 & -      \\
    \bottomrule
    \end{tabular}%
    }
    \caption{Pairwise signed dominance indices for the NDX dataset.
Each entry \((i,j)\) reports
\(\mathcal{D}(X^{(i)},X^{(j)})\).}
  \label{tab:pairwise_dominance_NDX}%
\end{table}%

\clearpage

{\footnotesize
\bibliographystyle{apalike}
\bibliography{References}
}

\end{document}